\documentclass{article}

\usepackage[english]{babel}
\usepackage{tabularx}
\usepackage{booktabs}
\usepackage{multirow}
\usepackage{amsmath}
\usepackage{bm}
\usepackage{CJKutf8}
\usepackage{diagbox} 
\usepackage{float}
\usepackage{amssymb}
\usepackage{authblk} 
\usepackage[letterpaper,top=2cm,bottom=2cm,left=3cm,right=3cm,marginparwidth=1.75cm]{geometry}
\usepackage{amsmath}
\usepackage{graphicx}
\usepackage[colorlinks=true, allcolors=blue]{hyperref}

\newcounter{rownumber}

\title{\textbf{Learning step-level dynamic soaring in shear flow}}

\author[1]{\fontsize{11pt}{14pt}\selectfont{Lunbing Chen}}
\author[1]{{Jixin Lu}}
\author[1]{{Yufei Yin}}
\author[1]{{Jinpeng Huang}}
\author[1,*]{{Yang Xiang}}
\author[1,\dag]{{Hong Liu}}

\affil[ ]{* \href{mailto:xiangyang@sjtu.edu.cn}{xiangyang@sjtu.edu.cn} (Lead Contact), \textsuperscript{\dag} \href{mailto:hongliu@sjtu.edu.cn}{hongliu@sjtu.edu.cn}}

\affil[1]{J.C.Wu Center for Aerodynamics, School of Aeronautics and Astronautics, Shanghai Jiao Tong University, Shanghai 200240, PR China}

\begin{document}
\begin{CJK}{UTF8}{gbsn}

\date{}
\maketitle









\begin{abstract}
Dynamic soaring enables sustained flight by extracting energy from wind shear, yet it is commonly understood as a cycle-level maneuver that assumes stable flow conditions. In realistic unsteady environments, however, such assumptions are often violated, raising the question of whether explicit cycle-level planning is necessary.
Here, we show that dynamic soaring can emerge from step-level, state-feedback control using only local sensing, without explicit trajectory planning. Using deep reinforcement learning as a tool, we obtain policies that achieve robust omnidirectional navigation across diverse shear-flow conditions.
The learned behavior organizes into a structured control law that coordinates turning and vertical motion, giving rise to a two-phase strategy governed by a trade-off between energy extraction and directional progress. The resulting policy generalizes across varying conditions and reproduces key features observed in biological flight and optimal-control solutions.
These findings identify a feedback-based control structure underlying dynamic soaring, demonstrating that efficient energy-harvesting flight can emerge from local interactions with the flow without explicit planning, and providing insights for biological flight and autonomous systems in complex, flow-coupled environments.
\end{abstract}


\section{Introduction}
\label{sec:intro}

Dynamic soaring (DS) is a flight strategy that enables seabirds, most notably albatrosses, to travel thousands of kilometers over the ocean by extracting energy from atmospheric wind shear \cite{weimerskirch2000fast, weimerskirch2002gps, goto2017asymmetry, kempton2022optimization, rayleigh1883soaring, sachs2013experimental, mohamed2022opportunistic}. This energy-harvesting mechanism represents an extreme form of efficient locomotion and has inspired the development of long-endurance autonomous aerial systems \cite{mohamed2022opportunistic, langelaan2009enabling}.

Existing studies on dynamic soaring span biological observations \cite{weimerskirch2000fast, weimerskirch2002gps, goto2017asymmetry, kempton2022optimization, sachs2013experimental, richardson2022observations, uesaka2023wandering, goto2024albatrosses}, trajectory optimization \cite{sachs2005minimum, deittert2009engineless, chen2025optimal}, reduced-order modeling \cite{taylor2016soaring, bousquet2017optimal, sachs2019kinetic, bousquet2017dynamic}, and control design \cite{harms2025robust, kai2019novel}. Despite their diversity, most approaches adopt a trajectory-level or cycle-level description, in which energy extraction is characterized over complete soaring maneuvers between wind layers \cite{sachs2013experimental, goto2024albatrosses, bousquet2017optimal}. These formulations implicitly assume that the flow remains sufficiently stable over each maneuver, enabling planning over an entire cycle.

In realistic unsteady environments, however, wind fields are highly variable and spatially heterogeneous \cite{langelaan2012wind, jones2022physics}. Flow conditions can change on spatial and temporal scales comparable to a single maneuver, violating the assumptions underlying cycle-level descriptions. As a result, predefined cyclic trajectories may become suboptimal, dynamically infeasible, or fail altogether when the flow deviates from assumed structures  \cite{harms2025robust, bird2014closing, hong2023dynamic, bronz2021flight}. This discrepancy challenges the view of dynamic soaring as a planning problem over fixed trajectories, and instead suggests that effective behavior may rely on step-level control based on instantaneous state information.

Achieving such a step-level description is fundamentally challenging \cite{deittert2009engineless, bird2014closing}. The agent must operate in a high-dimensional, nonlinear, and stochastic environment, relying only on local observations while achieving long-range navigation through sustained energy extraction \cite{reddy2016learning, gunnarson2021learning, jiao2025sensing, notter2023hierarchical, adamski2023towards}. Moreover, dynamic soaring couples two competing objectives: harvesting energy from the wind shear and maintaining directional progress toward a navigation goal \cite{weimerskirch2000fast, goto2024albatrosses, chen2025optimal}.
This leads to a central question: \textit{Is explicit cycle-level global planning necessary for dynamic soaring, or can sustained energy extraction and navigation emerge from step-level feedback based solely on local sensing?}

Recent advances in deep reinforcement learning (DRL) provide a potential framework for addressing this question \cite{reddy2016learning, gunnarson2021learning, jiao2025sensing, flato2024revealing, darveniza2026larval}. Unlike trajectory optimization, DRL learns closed-loop policies through interaction with the environment and can capture state-dependent feedback under stochastic and partially observed conditions. DRL has been successfully applied to dynamic soaring and related tasks \cite{montella2014reinforcement, verma2018efficient, abozeid2023comprehensive, dipaola2023framework, akhtar2026dynamic}. 
However, most existing studies use DRL primarily for trajectory generation or performance optimization, thereby retaining a trajectory-centric perspective and leaving unresolved whether dynamic soaring fundamentally requires planning or can emerge from feedback.

In this work, we formulate dynamic soaring as a closed-loop navigation problem and use DRL as a scientific tool to uncover its control structure. We show that dynamic soaring does not require explicit cycle-level planning, but can instead emerge from \textbf{step-level, state-feedback control using only local sensing}. The learned policies exhibit robust omnidirectional navigation in both uniform and spatially varying shear flows.
Analysis of the learned behavior reveals that dynamic soaring organizes into a structured control law. These findings identify a feedback-based control structure underlying dynamic soaring, demonstrating that efficient energy-harvesting flight can emerge from local interactions with the flow without explicit planning. This perspective reframes dynamic soaring as a feedback-driven control process and provides a principled foundation for understanding biological flight and designing autonomous systems for energy-efficient navigation in complex wind environments.

\section{Results}
\label{sec:results}

\subsection{RL achieves step-level dynamic soaring navigation}
\label{subsec:rl_framework}

\begin{figure}[htbp]
    \centering
    \includegraphics{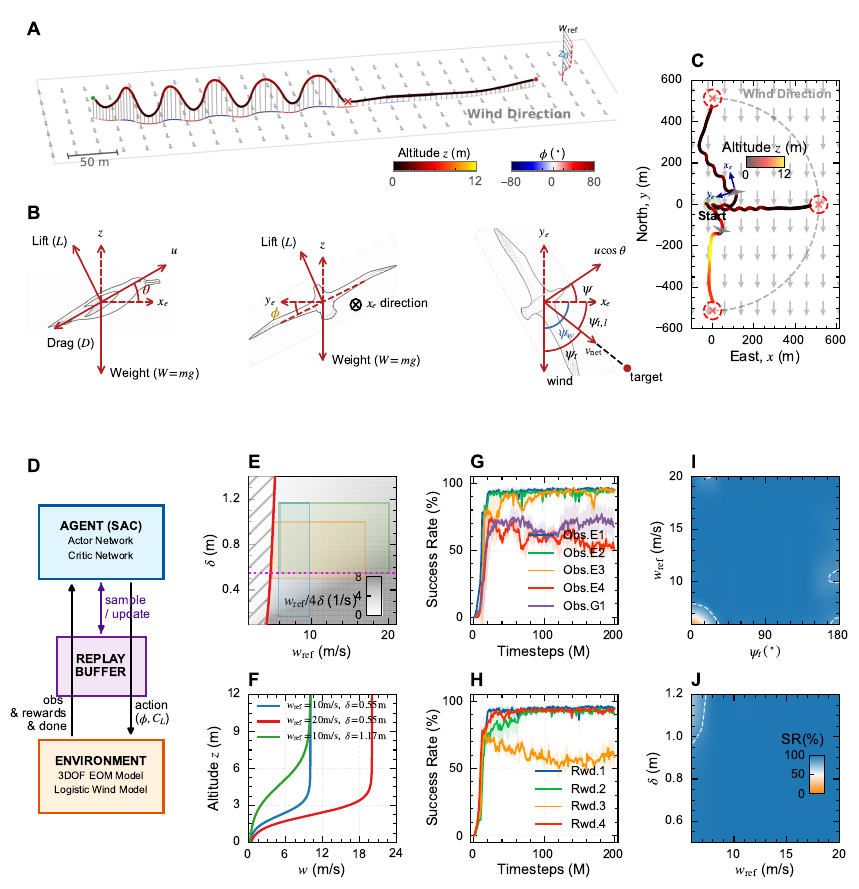}
    \caption{
    \textbf{Problem formulation and deep reinforcement learning framework for autonomous dynamic soaring.} 
    \textbf{(A)} Three-dimensional trajectory of the navigation task. 
    \textbf{(B)} The point-mass glider model \cite{chen2025optimal}. The egocentric frame $(x_e,y_e,z)$ denotes heading, left-wing, and up directions. $\mathbf{u}$, $\mathbf{v}$, and $\mathbf{w}$ represent airspeed, ground velocity, and wind velocity. The aerodynamic states are defined by pitch $\theta$, heading $\psi$, and bank angle $\phi$ \cite{anderson2011ebook}. $L$, $D$, and $W$ denote lift, drag, and weight. 
    $\psi_w$ and $\psi_t$ denote the wind direction and the target direction relative to the wind in the inertial (Earth-fixed) frame.
    \textbf{(C)} Horizontal projections of representative tailwind, crosswind, and headwind trajectories, illustrating the characteristic zig-zag motion \cite{goto2024albatrosses}. The circular regions indicate the randomized initial positions and the target success regions.
    \textbf{(D)} Deep reinforcement learning (DRL) framework. The agent interacts with the wind environment through observations and outputs continuous control actions $(\phi, C_L)$.
    \textbf{(E)} Parameter space of wind conditions. Grey shading indicates available energy ($\sigma_\mathrm{max}=w_{\mathrm{ref}}/(4\delta)$, \autoref{subsec:model}). Shaded regions denote prior studies: blue \cite{bousquet2017optimal}, yellow \cite{goto2022did}, and green this work. The red line marks the minimum wind-speed boundary \cite{bousquet2017dynamic} (left region infeasible), and the magenta dotted line indicates the theoretical perceptual resolution limit $l_{\mathrm{decision}} = v_c t_{\mathrm{decision}}$.
    \textbf{(F)} Logistic wind profiles.
    \textbf{(G, H)} Training success-rate curves under different observation settings (\autoref{tab:obs_ablation}) and reward formulations (\autoref{tab:reward_ablation}). Curves are averaged over five random seeds, with shaded regions indicating standard deviation.
    \textbf{(I, J)} Success rate as a function of $(\psi_t,\, w_{\mathrm{ref}})$, and  $(w_{\mathrm{ref}}, \, \delta)$, where white dashed contours denote the $90\%$ SR boundary. Statistical definitions of success are provided in \autoref{para:SR}.
    }
    \label{fig:problem}
\end{figure}

We formulated dynamic soaring as a closed-loop navigation problem in a vertically shear wind field, and trained a model-free DRL agent to control a glider under diverse wind conditions (\autoref{fig:problem}A-D) \cite{dipaola2023framework}. 
The glider dynamics are represented by a six-dimensional state vector $\mathbf{s} = [u, \theta, \psi, x, y, z]^T$ (\autoref{fig:problem}B) \cite{chen2025optimal}. The wind field is modeled using a logistic profile (\autoref{fig:problem}A, E, F) \cite{bousquet2017optimal, goto2022did}, which captures the shear-layer structure associated with flow separation behind ocean waves more realistically than logarithmic \cite{sachs2005minimum, chen2025optimal} or linear models \cite{zhao2004optimal}.
At each time step, the agent receives a compact observation of its instantaneous flight state and local wind condition, and outputs continuous control commands (\autoref{fig:problem}D). The reward promotes sustained flight and directional progress while penalizing unstable or failed trajectories. Detailed model equations and training procedures are provided in \autoref{sec:methods}.

The navigation task formulation is designed to test whether robust dynamic soaring can emerge from local interaction with the flow, rather than from prescribing predefined maneuver cycles. The initial position $(x_0, y_0)$ is sampled within a circular region of radius $2l_c$, and a trial is considered successful if the agent reaches a target zone of the same radius (\autoref{fig:problem}C).
The task horizon is defined by a target distance $d_t = 600\, v_c t_{\mathrm{decision}}$, chosen to balance task difficulty and learnability. It exceeds the unpowered gliding range, requiring sustained energy harvesting, while remaining within the agent’s effective planning horizon. For a discount factor $\gamma = 0.995$, the effective horizon is $N_{\mathrm{eff}} \approx 920$, so that $600 < N_{\mathrm{eff}}$ ensures reliable propagation of the terminal reward to early states \cite{sutton1998reinforcement}.
To systematically evaluate navigation across wind-relative directions, the target directions relative to the wind $\psi_t$ are sampled in $[0^\circ, 180^\circ]$, spanning tailwind, crosswind, and headwind conditions. Owing to the bilateral symmetry of the system, the complementary angular range is redundant and is not explicitly trained.

Training curves are shown in \autoref{fig:problem}G, H. The success rate exceeds $95\%$ under Obs.E1 (\autoref{tab:obs_ablation}) and Rwd.1 (\autoref{tab:reward_ablation}). The agent remains airborne, continuously extracts energy from the shear layer, and achieves stable long-range navigation (\autoref{fig:problem}A, C).
The learned policy produces sustained dynamic-soaring trajectories across a wide range of conditions, maintaining high performance over diverse target directions ($\psi_t \in [0^\circ, 180^\circ]$), wind speeds ($w_\mathrm{ref} \in [6,20] \, \mathrm{m/s}$), and shear-layer thicknesses ($\delta \in [0.55, 1.17] \, \mathrm{m}$) (\autoref{fig:problem}I, J).
These results demonstrate that dynamic soaring can emerge from step-level, feedback-driven control using only local observations, without requiring explicit cycle-level planning.

\begin{table}[htbp]
    \centering
    \caption{
    \textbf{Observation ablation.}
    This table summarizes the observation configurations used in the sensory ablation study. All policies are trained with the same reward formulation (Rwd.1 in \autoref{tab:reward_ablation}). Success-rate definitions are provided in \autoref{para:SR}.
    E1 defines the full egocentric observation set. Removing shear information (E2) or replacing airspeed with groundspeed (E3) leads to moderate performance changes, while the polar representation (E4) significantly degrades learning. The geocentric formulation (G1) also shows reduced performance compared to the egocentric baseline.
    Configurations without wind (E0) or without airspeed information (E0') fail to learn meaningful policies. For some settings, curriculum learning does not cover the full $0^\circ$--$180^\circ$ range (\autoref{subsubsec:curriculum_learning}): E0 converges within $40^\circ$--$140^\circ$, E0' fails beyond $80^\circ$--$100^\circ$, and E4 converges within $10^\circ$--$170^\circ$. Reported success rates are computed over these respective ranges.
    }
    \label{tab:obs_ablation}
    \begin{tabular}{lllll}
        \toprule
        \textbf{No.} & \textbf{Setting} & \textbf{Obs.} &  \textbf{Training SR} & \textbf{Test SR}\\
        \midrule
        E1 & full  & \(\Delta x_e, \Delta y_e, z, u_{x,e}, u_{z}, w_{x,e}, w_{y,e}, \sigma_w\) & \(95.5\% \pm 0.7\%\) & \(97.3\% \pm 0.8\%\) \\
        E2 & No shear & \(\Delta x_e, \Delta y_e, z, u_{x,e}, u_{z}, w_{x,e}, w_{y,e}\) & \(93.8\% \pm 1.1\%\) & \(92.3\% \pm 1.8\%\) \\
        E3 & ground speed  & \(\Delta x_e, \Delta y_e, z, v_{x,e}, v_{y,e}, u_{z}, w_{x,e}, w_{y,e}, \sigma_w\) & \(93.1\% \pm 2.4\%\) & \(97.9\% \pm 0.7\%\) \\
        E4 & polar wind  & \(\Delta x_e, \Delta y_e, z, u_{x,e}, u_{z}, w, \psi_{w,e}, \sigma_w \) & \(53.7\% \pm 4.3\%\) &  \(62.4\% \pm 4.9\%\) \\
        G1 & geocentric & \(\Delta x_g, \Delta y_g, z, u_{x,g}, u_{y,g}, u_{z}, w_{x,g}, w_{y,g}, \sigma_w\) & \(74.8 \% \pm 8.0 \% \) & \(88.0\% \pm 4.4\% \) \\
        E0 & No wind & \(\Delta x_e, \Delta y_e, z, u_{x,e}, u_{z}\) & \(30.8 \% \pm 5.6 \%\) & \(23.0 \% \pm 9.0\%\) \\
        E0' & No airspeed & \(\Delta x_e, \Delta y_e, z, w_{x,e}, w_{y,e}\) & \(0.0\%\) & - \\
        \bottomrule
    \end{tabular}
\end{table}

\begin{table}[htbp]
    \centering
    \caption{
    \textbf{Process reward ablation.}
    Training and test success rates under different reward formulations. All policies use the same observation space (Obs.E1 in \autoref{tab:obs_ablation}). The combined reward (Rwd.1) achieves the best performance, while the directional-progress term alone (Rwd.3) yields comparable results. In contrast, the energy-based term alone (Rwd.2) fails to produce a robust policy. State-based rewards (Rwd.4) achieve moderate performance but remain inferior to process-based formulations.
    }
    \begin{tabular}{llll}
    \toprule
    \textbf{No.} & \textbf{\(r_\mathrm{process}\)} & \textbf{Training SR} & \textbf{Test SR}\\
    \midrule
    1 & \(\xi_{\dot{e}}\dot{e}/e_{\mathrm{norm}} \sigma_{\mathrm{norm}}
    + \xi_{v}\,v_{\mathrm{net}}/u\) & \(95.5\% \pm 0.7\%\) & \(97.3\% \pm 0.8\%\) \\
    2 & \(\xi_{v}\,v_{\mathrm{net}}/{u}\) & \( 92.6\% \pm  1.1\%\) & \( 94.5\% \pm 1.3\%\) \\
    3 & \(\xi_{\dot{e}}\dot{e}/e_{\mathrm{norm}} \sigma_{\mathrm{norm}}\) & \( 58.0\% \pm  3.3\%\) & \(62.4\% \pm 4.9\%\) \\
    4 & \(\xi_{e} \Delta e/e_{\mathrm{norm}}
    + \xi_{d} \Delta d/d_{\mathrm{norm}}\) & \( 93.4\% \pm  0.3\%\) & \( 91.8\% \pm  1.9\%\) \\
    \bottomrule
    \end{tabular}
    \label{tab:reward_ablation}
\end{table}

\subsection{Kinetic-energy-managed DS for long-range navigation}
\label{subsec:policy_macro}

\begin{figure}[htbp]
    \centering
    \includegraphics{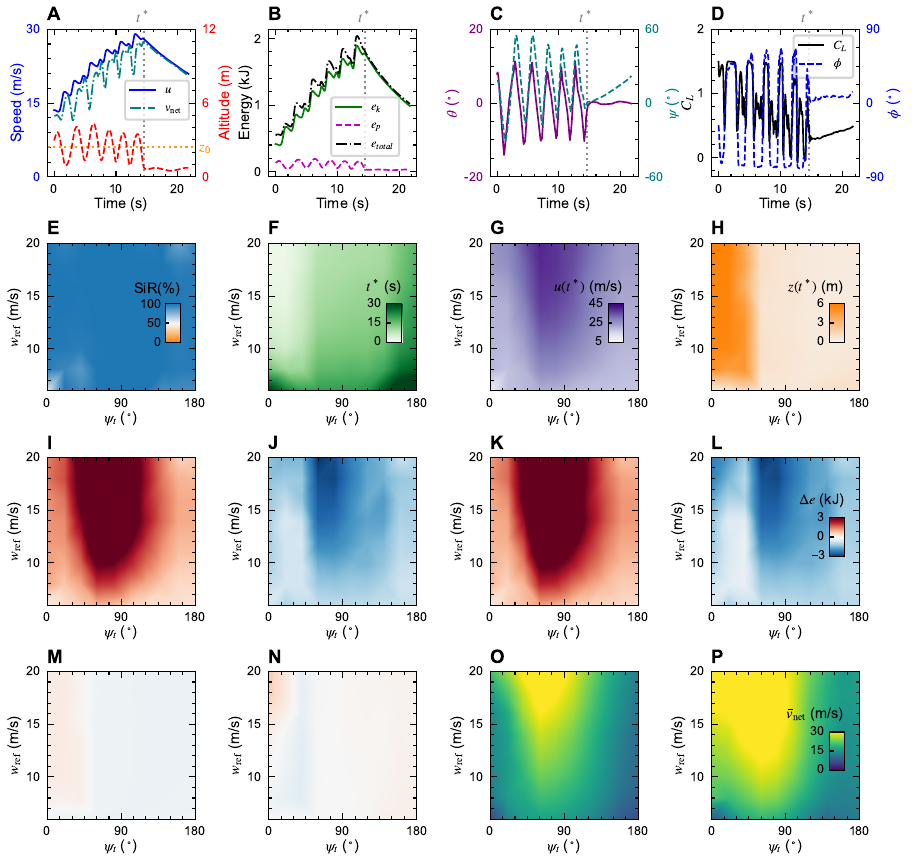}
    \caption{
    \textbf{Emergence of a two-phase dynamic-soaring navigation strategy governed by kinetic-energy management.}
    \textbf{(A–D)} Time evolution of key variables along a representative crosswind trajectory (\autoref{fig:problem}A): \textbf{(A)} airspeed $u$, ground-directed velocity $v_{\mathrm{net}}$, and altitude $z$; \textbf{(B)} total energy $e$, kinetic energy $e_k$, and potential energy $e_p$; \textbf{(C)} pitch angle $\theta$ and heading angle $\psi$; \textbf{(D)} control actions $C_L$ and $\phi$. The grey line indicates the transition time $t^*$ (\autoref{para:t_star}), separating the dynamic-soaring (DS) and targeted-gliding (TG) phases.
    \textbf{(E–H)} Statistical characterization across $(\psi_t, w_{\mathrm{ref}})$: \textbf{(E)} two-phase significance ratio (SiR, \autoref{para:SiR}); \textbf{(F)} transition time $t^*$ (\autoref{para:t_star}); \textbf{(G)} airspeed at transition $u(t^*)$; \textbf{(H)} altitude at transition $z(t^*)$.
    \textbf{(I–P)} Energy and navigation performance in the two phases. \textbf{(I–J)} net change in total energy $\Delta e=\Delta e_k+\Delta e_p$; \textbf{(K–L)} net change in kinetic energy $\Delta e_k$; \textbf{(M–N)} net change in potential energy $\Delta e_p$; \textbf{(O–P)} mean directional velocity $\bar{v}_{\mathrm{net}}$. Panels \textbf{(I, K, M, O)} correspond to the DS phase, and \textbf{(J, L, N, P)} to the TG phase. Energy changes are computed relative to the transition time $t^*$: $\Delta e_{\mathrm{DS}}=e(t^*)-e(0)$ and $\Delta e_{\mathrm{TG}}=e(t_N)-e(t^*)$, with analogous definitions for $\Delta e_k$ and $\Delta e_p$. $\bar{v}_{\mathrm{net}}$ is averaged over each phase.
    These results show that kinetic energy is accumulated during the DS phase and expended during the TG phase to enable goal-directed navigation.
    }
    \label{fig:policy_macro}
\end{figure}

The learned policy exhibits a robust two-phase structure for long-range navigation, consisting of a dynamic soaring (DS) phase followed by a targeted gliding (TG) phase. As shown in \autoref{fig:problem}A, C, representative trajectories initially display a periodic zig-zag motion characteristic of dynamic soaring, and subsequently transition to a near-straight glide toward the target. The associated state variables (\autoref{fig:policy_macro}A-D) show consistent behavior: oscillatory dynamics during the DS phase followed by smooth, monotonic evolution during the TG phase.

This behavior can be understood as a process of kinetic-energy management. During the DS phase, the agent repeatedly traverses the shear layer (\autoref{fig:problem}A, \autoref{fig:policy_macro}A) and accumulates kinetic energy through interaction with the wind gradient (\autoref{fig:policy_macro}B) \cite{kempton2022optimization, sachs2012flying}, leading to oscillatory but overall increasing energy levels. In contrast, during the TG phase, the agent exits the shear region and gradually converts the stored kinetic energy into forward motion toward the target (\autoref{fig:problem}A, \autoref{fig:policy_macro}A, B). Quantitatively, the variation in kinetic energy dominates that of potential energy ($\Delta e_k \sim O(10^3)$ versus $\Delta e_p \sim O(10^2)$, \autoref{fig:policy_macro}B, I-N), indicating that successful navigation is governed primarily by kinetic-energy acquisition and expenditure rather than altitude-based potential energy storage. Consistent with this interpretation, the net ground-directed velocity $v_\mathrm{net}$ remains relatively low during the DS phase, reflecting the effort of energy harvest (\autoref{fig:policy_macro}A, O--P).

The two-phase structure remains robust across stochastic conditions, with target direction ($\psi_t\in[0^\circ,180^\circ]$), wind speed ($w_{\mathrm{ref}}\in[6,20]\,\mathrm{m/s}$), and shear thickness ($\delta\in[0.55,1.17]\,\mathrm{m}$) sampled over broad ranges.  Representative trajectories are shown in \autoref{fig:S_fig02_1}. Over $96\%$ of trajectories display statistically distinguishable phases (\autoref{fig:policy_macro}E, \autoref{fig:S_fig02_2}A), demonstrating that this macro strategy emerges as a general solution rather than a condition-specific behavior. Deviations occur primarily under weak-wind or thick-shear conditions, where reduced energy availability and lower success rates obscure the phase distinction (\autoref{fig:problem}I-J).

While the DS--TG strategy is consistent, its detailed manifestation depends on environmental conditions (\autoref{fig:policy_macro}F-P). In particular, the transition between phases is modulated by the target direction relative to the wind. For downwind targets ($\psi_t \lesssim 60^\circ$), the agent typically transitions above the shear layer (\(z(t^*) > z_0\), \autoref{fig:policy_macro}H), exploiting high-speed free-stream flow for efficient gliding ($v \approx u+w$). In contrast, for crosswind and upwind targets ($\psi_t \gtrsim 60^\circ$), the transition occurs below the shear layer (\(z(t^*) < z_0\)), where reduced wind speeds mitigate drift ($v \approx u-w$) and improve directional control \cite{richardson2018flight}. These differences also affect transition time and airspeed. Since the wind component aligned with the target direction directly increases the directional velocity, downwind navigation transitions earlier (\autoref{fig:policy_macro}F) and requires less airspeed accumulation (\autoref{fig:policy_macro}G).
Variations in wind strength and shear thickness primarily influence the magnitude of available energy, while preserving the underlying two-phase structure (\autoref{fig:S_fig02_2}).

The emergence of the DS--TG structure can be understood as the result of the interaction between reinforcement-learning objectives and physical constraints. The discounted reward formulation encourages the agent to reach the target as early as possible (\autoref{subsubsec:rwd_design}), favoring transitions to energetically efficient, goal-directed motion once sufficient energy has been accumulated. At the same time, physical and aerodynamic constraints (\autoref{subsubsec:rwd_design}) limit unbounded energy growth during dynamic soaring. As a result, the agent naturally adopts a strategy in which energy is first accumulated through dynamic soaring and then expended through efficient gliding.

\subsection{Structured step-level state-feedback control law for DS}
\label{subsec:policy_micro}

The learned policy defines a structured state-feedback control law, in which control actions are determined by local wind and kinematic states.

\begin{figure}[htbp]
    \centering
    \includegraphics{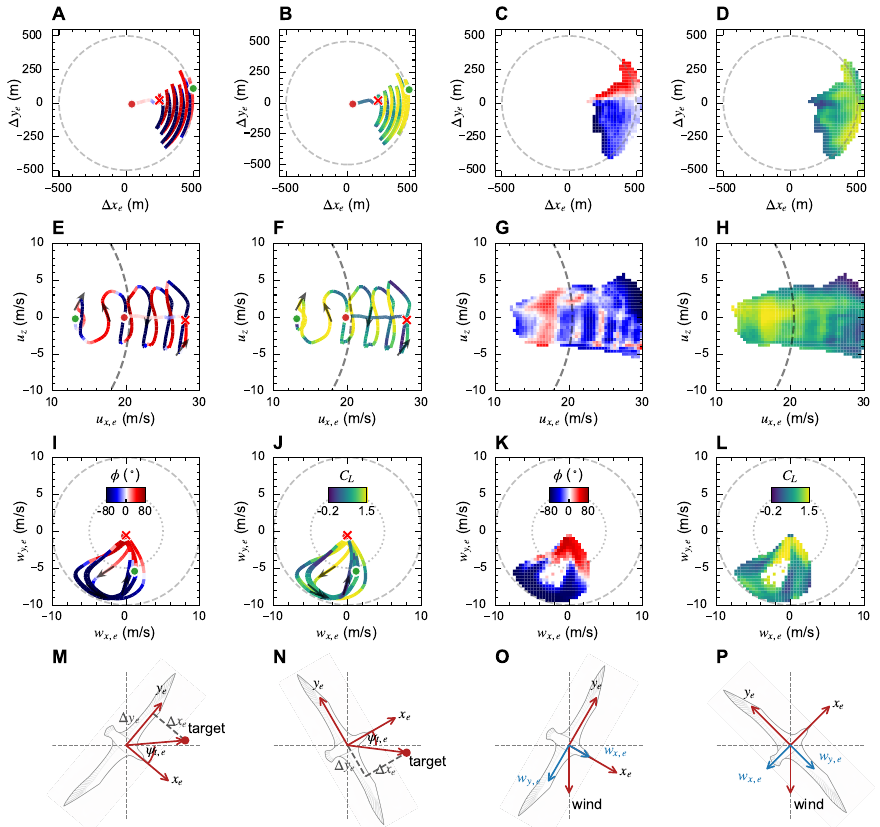}
    \caption{
    \textbf{Structured policy representation in observation space under a fixed condition} ($\psi_t = 90^\circ$, $w_\mathrm{ref}=10\,\mathrm{m/s}$, $\delta = 0.55\,\mathrm{m}$).
    \textbf{Columns 1–2} show a representative successful trajectory colored by $\phi$ and $C_L$, with the start, target, and DS--TG transition marked by a green circle, red circle, and red cross. \textbf{Columns 3–4} show occupancy-filtered heatmaps from $1,000$ successful DS-phase trajectories (TG phase in \autoref{fig:S_fig03_1}), retaining states with more than $100$ samples.
    \textbf{Rows 1–3} correspond to relative position $(\Delta x_e,\Delta y_e)$, velocity $(u_{x,e},u_z)$, and relative wind $(w_{x,e},w_{y,e})$. Trajectories show temporal evolution, while heatmaps reveal consistent observation–action mappings.
    In $(\Delta x_e, \Delta y_e)$, the state encodes distance $d_e$ and target direction $\psi_{t,e}$, organizing navigation and the DS–TG transition. In $(u_{x,e},u_z)$, the auxiliary curve indicates the $3g$ load-factor limit for $u=u_c$. In $(w_{x,e},w_{y,e})$, concentric circles at $5$ and $10\,\mathrm{m/s}$ indicate characteristic shear-layer and free-stream wind magnitudes.
    The final row shows coordinate mappings: \textbf{(M, N)} $(x_e,y_e)\rightarrow (d_e,\psi_{t,e})$ and \textbf{(O, P)} $(w_{x,e},w_{y,e})\rightarrow (w,\psi_{w,e})$.
    }
    \label{fig:policy_micro}
\end{figure}

The observation spaces used here provide an interpretable view of the policy.
The egocentric position $(\Delta x_e, \Delta y_e)$ specifies the relative target direction and distance, providing the geometric reference for navigation and the DS--TG phase transition (\autoref{fig:policy_micro}A-D, M-N). In the DS phase (\autoref{fig:policy_micro}C-D), trajectories occupy a broad sector in this space, whereas in the TG phase (\autoref{fig:S_fig03_1}A-B) they collapse toward $\Delta y_e \approx 0$, indicating alignment with the target. The velocity state $(u_{x,e}, u_z)$ encodes airspeed and vertical motion, reflecting both aerodynamic feasibility and the current kinetic-energy level (\autoref{fig:policy_micro}E-H). The wind state $(w_{x,e}, w_{y,e})$ encodes local flow conditions: its magnitude reflects the position of the agent relative to the shear layer and thus the available environmental energy, while its direction specifies the relative orientation between the agent and the flow (\autoref{fig:policy_micro}I-L, O-P). Together, these state variables make the learned control structure directly observable.

The bank angle $\phi$ regulates horizontal reorientation as a function of the wind-relative state. According to the heading-rate relationship ($\dot{\psi} \propto \sin\phi$, \autoref{eq:EOM_dot_psi}), the sign of $\phi$ determines the turning direction. During the DS phase (\autoref{fig:policy_micro}I, K), $\phi$ exhibits a structured dependence on the wind state: large magnitudes appear in both low- and high-wind regions, indicating active turning, while $\phi \approx 0$ near the shear-layer center corresponds to near-straight motion. The sign of $\phi$ encodes directional decisions (\autoref{fig:policy_micro}K, O, P): in low-wind regions, $\phi>0$ induces upwind turning ($\psi$ increasing), whereas in high-wind regions, $\phi<0$ produces downwind turning. This establishes a direct mapping from wind state to horizontal control. During the TG phase (\autoref{fig:S_fig03_1}), $\phi \approx 0$, corresponding to near-straight flight toward the target.

The lift coefficient $C_L$ governs vertical motion as a state-dependent control input ($\dot{\theta} \propto C_L$, \autoref{eq:EOM_dot_theta}). During the DS phase (\autoref{fig:policy_micro}J, L), $C_L$ depends primarily on the wind state: larger values are selected in low-wind regions to induce ascent, whereas reduced values in high-wind regions produce descent, generating the alternating climb--descent pattern required for sustained DS cycles. This control is further modulated by airspeed. As the airspeed increases (\autoref{fig:policy_micro}F, H), the admissible range of large $C_L$ values is restricted by the load-factor constraint ($n \propto u^2$, \autoref{subsubsec:rwd_design}), leading to a narrowing of the feasible control range. During the TG phase (\autoref{fig:S_fig03_1}), $C_L$ varies smoothly to maintain approximately level gliding as the airspeed decreases.

Taken together, these results reveal a structured state-feedback control law in which $\phi$ and $C_L$ are jointly determined by wind and kinematic states to regulate horizontal turning and vertical motion. This produces a consistent four-stage sequence: upwind turning in low-wind regions, near-straight climbing across the shear layer, downwind turning in high-wind regions, and near-straight descent back into the low-wind region (\autoref{fig:problem}A). This sequence corresponds to the canonical dynamic-soaring pattern of \emph{ascending upwind and descending downwind} \cite{taylor2016soaring}. Importantly, this structure is not imposed but emerges from the learned policy, indicating that dynamic soaring can be understood as a physics-consistent control law derived from local state feedback. Furthermore, this policy remains consistent across different training checkpoints (\autoref{fig:S_fig03_2}) and under varying target directions (\autoref{fig:S_fig03_3}) and wind conditions (\autoref{fig:S_fig03_4}).


\subsection{Wind-relative sensing for DS control}
\label{subsec:observations}

To identify the sensory information underlying the learned control policy, a systematic observation ablation is performed across stochastic navigation tasks and wind conditions, with $\psi_t \in [0^\circ, 180^\circ]$, $w_{\mathrm{ref}} \in [6, 20]\,\mathrm{m/s}$, and $\delta \in [0.55, 1.17]\,\mathrm{m}$.
Detailed observation design is provided in \autoref{subsubsec:obs_design}. These results allow us to relate sensing structure to the state-feedback control law identified in \autoref{subsec:policy_micro}.

Relative representation enables consistent control.
A wind-relative (egocentric) representation is critical for both robust control and generalization. As shown in \autoref{tab:obs_ablation} and \autoref{fig:S_fig01_1}, egocentric observations achieve test success rates above $95\%$, whereas geocentric observations remain below $90\%$. Under varying wind directions, geocentric policies fail to transfer, with success rates dropping to $0\%$ when $\psi_w$ deviates from the training configuration, while egocentric policies maintain success rates above $99\%$ (\autoref{fig:S_fig01_2}). These results indicate that the learned control law relies on invariant geometric relationships between the agent, the target, and the flow, which are naturally preserved in a relative frame \cite{jiao2025sensing}.

Flow-gradient information resolves control ambiguity.
Including explicit shear information improves performance, particularly in low-environment-energy conditions. Observation sets that include the vertical wind gradient consistently outperform those based on wind speed alone (\autoref{tab:obs_ablation}, \autoref{fig:S_fig01_1}). The difference becomes most pronounced in weak-wind or thick-shear regimes (\autoref{fig:S_fig01_1}I), where the available energy is limited \cite{richardson2022observations, chen2025optimal}. Without gradient information, identical wind speeds may correspond to different positions within the shear layer \cite{stull2012introduction}, rendering such states indistinguishable and leading to ambiguous control decisions. Providing shear information resolves this ambiguity and supports consistent state-dependent control.

Airspeed sensing supports stable and feasible control.
Although airspeed- and groundspeed-based observations are mathematically equivalent (\autoref{subsubsec:obs_design}) and yield similar success rates (\autoref{tab:obs_ablation}), their training dynamics differ significantly (\autoref{fig:S_fig01_1}). Groundspeed-based policies exhibit slower convergence and repeated performance collapses (e.g., around 70M and 170M steps), indicating unstable learning dynamics. In contrast, airspeed-based sensing provides direct access to aerodynamic state variables, enabling stable regulation of lift and improved robustness during training.

Representation structure affects learnability.
Despite containing equivalent information, Cartesian wind components enable reliable learning, whereas magnitude--angle representations fail to converge (\autoref{tab:obs_ablation}). This suggests that representations aligned with the underlying flight dynamics are easier for the policy to exploit \cite{gunnarson2021learning, jiao2025sensing}, while polar forms introduce additional nonlinearities that hinder learning.

Together, these results show that effective dynamic-soaring control relies on a compact wind-relative sensing structure that encodes flow orientation, shear variation, and aerodynamic state. This sensing configuration aligns with the control dependencies identified in \autoref{subsec:policy_micro}, where wind-related states govern directional control and airspeed constrains vertical maneuvering.


\subsection{DS is a multi-objective process}
\label{subsec:reward}

Dynamic soaring navigation is inherently a multi-objective process, in which the agent must balance energy acquisition and directional progress toward the target \cite{goto2024albatrosses, chen2025optimal}. Using reward ablation in a DRL framework \cite{flato2024revealing, stuber2025miniature}, we examine this trade-off directly at the control level (reward design in \autoref{subsubsec:rwd_design}).

Process-based rewards are necessary for stable and robust learning. As shown in \autoref{tab:reward_ablation} and \autoref{fig:S_fig01_4}, policies trained with state-based rewards fail under challenging environmental conditions, particularly in weak-wind and thick-shear regimes (\autoref{fig:S_fig01_4}L). In contrast, process-based rewards, which provide direct guidance on flight evolution, yield consistently higher success rates and more stable control behavior.

Within this formulation, directional progress is the dominant objective. A reward based solely on $v_{\mathrm{net}}$ achieves nearly the same performance as the full formulation, whereas a reward based on $\dot{e}$ alone fails to produce a successful policy (\autoref{tab:reward_ablation}). Moreover, in the combined formulation, the contribution of the $\dot{e}$ term remains secondary compared to the directional term (\autoref{fig:S_fig01_5}). This indicates that explicit directional guidance is essential for navigation.

Energy acquisition, by contrast, emerges implicitly through survival constraints. Even without $\dot{e}$, the crash penalty enforces a minimum energy level required to remain airborne. Training dynamics support this interpretation (\autoref{fig:S_fig01_4}A-D): the agent first learns to avoid crashes and extend survival time, before improving directional efficiency. 

Together, these results show that dynamic soaring control is governed by a trade-off between energy acquisition and directional progress. Energy-related objectives primarily enhance robustness, whereas direction-related objectives ensure successful navigation, indicating that effective strategies lie along a Pareto frontier between these competing objectives \cite{shoval2012evolutionary}.






\section{Discussions}
\label{sec:discussions}

\subsection{Generalization to unseen conditions}
\label{subsec:generalization}

To assess whether the learned policy captures transferable physical principles rather than overfitting to the training distribution \cite{sutton1998reinforcement, lillicrap2020continuous, peng2018sim}, its performance is evaluated under three categories of out-of-distribution conditions: spatially varying wind fields, altered navigation tasks, and noisy observations. The generalization setup is detailed in \autoref{subsec:gen_setup}.

\begin{figure}[htbp]
    \centering
    \includegraphics{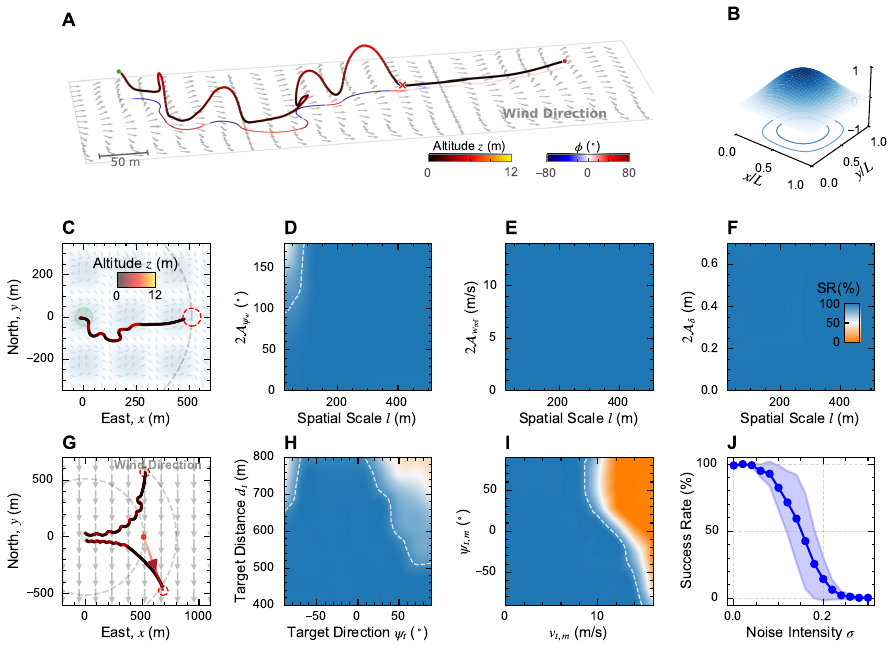}
    \caption{
    \textbf{Robustness and generalization under out-of-distribution conditions.}
    \textbf{(A, C)} Representative trajectory in a spatially varying wind field with coupled speed and shear variations.
    \textbf{(B)} Normalized spatial distribution of the harmonic disturbance field $H(\mathbf{p})$ (defined in \autoref{subsec:gen_setup}).
    \textbf{(D–F)} Success-rate heatmaps under perturbed wind conditions, showing robust performance across variations in wind-direction scale $\psi_w$, reference wind speed $w_{\mathrm{ref}}$, and shear thickness $\delta$.
    \textbf{G} Representative trajectories for navigation tasks.
    \textbf{(H–I)} Success-rate heatmaps for static targets $(\psi_t, d_t)$ and dynamic targets $(v_{t,m},\psi_{t,m})$.
    \textbf{(J)} Robustness to observation noise with increasing intensity $\sigma$.
    The policy maintains high performance across environmental variability, task complexity, and sensing uncertainty.
    }
    \label{fig:generalization}
\end{figure}

The policy maintains success rates above $95\%$ under spatially varying wind environments (\autoref{fig:generalization}A–F), despite being trained only in uniform wind fields. This strong performance indicates that the agent exploits local wind-gradient information rather than memorizing fixed trajectories. Performance degrades only when the spatial variation occurs at sufficiently small length scales.
This failure arises from physical maneuverability limits rather than a lack of policy generalization. Assuming that the lateral component of lift provides centripetal acceleration ($L\sin\phi = m u^2 / R$), the turning radius is constrained by the balance between aerodynamic force and inertial motion. This yields a minimum turning length scale of $l_{\min} = \pi R \approx 2\pi m / (\rho S C_L \sin\phi) \approx 87\,\mathrm{m}$. 
This scale closely matches the boundary of degraded performance observed in \autoref{fig:generalization}C. When flow variations occur below this scale, they exceed the agent’s reorientation capability, leading to reduced success rates.

The policy also generalizes to navigation tasks beyond the training setting (\autoref{fig:generalization}G, H). For static targets, the distance $d_t$ is varied from $300$ to $800\,\mathrm{m}$, and performance degrades at large distances, primarily due to observation extrapolation beyond the training distribution, leading to timeout rather than crash failures (\autoref{fig:S_fig04_1}). Notably, the agent remains airborne in these cases, indicating that energy-harvesting behavior is preserved even when directional guidance fails (\autoref{fig:S_fig04_1}I-L). 

For dynamic targets (\autoref{fig:generalization}G, I), the agent successfully tracks moving goals across a wide range of velocities and directions. In challenging scenarios, particularly under strong headwind conditions, failures are again dominated by timeout rather than crash. Trajectory analysis (\autoref{fig:S_fig04_2}I-L) shows that the agent can re-enter dynamic-soaring phases after initiating a glide, demonstrating adaptive re-planning behavior. This ability to switch between DS and TG modes in response to task demands indicates that the learned policy encodes a reusable control strategy.

The policy remains stable under observation noise. As shown in \autoref{fig:generalization}J, performance is maintained for noise levels up to $10\%$ of the observation magnitude. This robustness indicates that the controller operates as a closed-loop feedback system rather than relying on precise state estimation \cite{todorov2002optimal}. The neural policy directly maps noisy observations to consistent actions, effectively learning implicit noise filtering and stabilization.

Across all tests, the policy exhibits consistent behavior: it adapts to environmental variation, maintains dynamic-soaring dynamics under task perturbations, and remains stable under noisy observations. These results indicate that the agent has learned a generalizable state-feedback control law grounded in the physics of wind-gradient exploitation, rather than a task-specific trajectory.

\subsection{Comparison with biological data and optimal control}
\label{subsec:comparison}

The learned policy is both biologically consistent and near-optimal. It reproduces key features of animal flight while approaching the performance of optimal-control solutions.

\begin{figure}[htbp]
    \centering
    \includegraphics{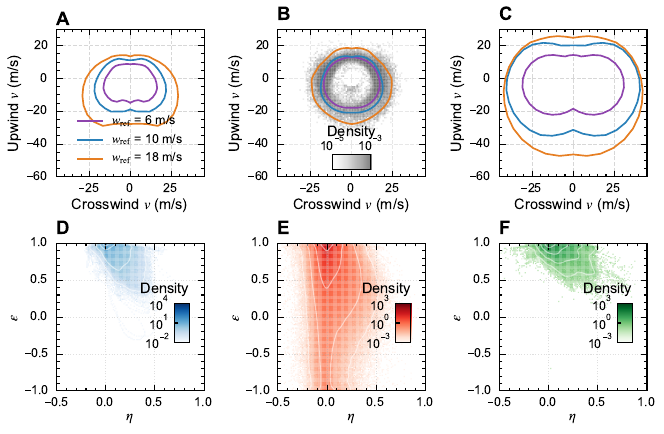}
    \caption{
    \textbf{Comparison of ground-speed envelopes and energy-direction trade-offs across learned, biological, and optimal strategies.}
    \textbf{(A–C)} Ground-speed envelopes under different wind conditions. (A) RL policy predictions for $w_\mathrm{ref}=6, 10, 18\,\mathrm{m/s}$ in polar coordinates. (B) Experimental envelopes derived from biological flight data \cite{uesaka2023wandering}, fitted using a generalized additive model \cite{goto2024albatrosses}, with background shading indicating data density. (C) Numerically optimal envelopes obtained via IPOPT-based trajectory optimization \cite{chen2025optimal}.
    \textbf{(D–F)} Joint distributions of energy-harvesting coefficient $\eta$ and directional-flight coefficient $\epsilon$ (defined in \cite{chen2025optimal}) for $\psi_t \in [60^\circ,120^\circ]$. Panels show (D) RL policy samples, (E) filtered experimental data, and (F) IPOPT solutions. Color maps indicate log-scaled density with overlaid contours.
    All three approaches exhibit a consistent trade-off structure between energy acquisition and directional progress, despite variability in experimental data.
    }
    \label{fig:comparision}
\end{figure}

The learned policy captures the wind-dependent structure of ground-speed distributions observed in nature \cite{goto2017asymmetry}. As shown in \autoref{fig:comparision}A–C, it reproduces the characteristic ``butterfly-shaped'' pattern reported in biological data \cite{goto2024albatrosses, richardson2018flight}. Compared to IPOPT-based optimal solutions \cite{chen2025optimal}, the RL policy more closely matches experimentally observed trends. Minor discrepancies at high wind speeds (e.g., $w_{\mathrm{ref}}\approx 18\,\mathrm{m/s}$) are likely due to sparse experimental sampling, whereas agreement at moderate wind speeds ($w_{\mathrm{ref}}=6,10\,\mathrm{m/s}$) is strong.

The learned policy also reproduces the fundamental trade-off between energy acquisition and directional flight. As shown in \autoref{fig:comparision}D–F, both the RL policy and IPOPT solutions exhibit a clear trade-off structure, with $\epsilon$ decreasing as $\eta$ increases, consistent with theoretical predictions \cite{chen2025optimal}. Experimental data show the same trend, with probability mass shifting toward higher $\eta$ and lower $\epsilon$ \cite{yonehara2016flight}. Occasional cases with $\epsilon \approx 0$ correspond to backward or reversing segments in measured trajectories (\autoref{fig:S_fig05_1}), which are not present in RL or optimal-control solutions but do not alter the overall trade-off structure.


\subsection{Conclusion and future work}
\label{subsec:conclusion}

In this study, we show that dynamic soaring does not require explicit cycle-level planning, but can instead emerge from step-level, state-feedback control using only local sensing. The learned policies achieve robust omnidirectional navigation ($\psi_t \in [0^\circ,180^\circ]$) across a wide range of wind conditions ($w_\mathrm{ref} \in [6,20] \, \mathrm{m/s}, \, \delta \in [0.55, 1.17] \, \mathrm{m}$) and reveal a consistent underlying control structure.

Our results identify three key elements of this feedback-based strategy. First, dynamic soaring can be described as a reusable step-level control law operating on instantaneous state information. Second, effective control relies on a compact wind-relative sensing representation that captures the essential flow geometry. Third, long-range navigation is governed by a fundamental trade-off between energy harvesting and directional progress. Together, these findings provide a unified interpretation of dynamic soaring as a feedback-driven control process.

This perspective reframes dynamic soaring from a trajectory planning problem to a feedback control problem in flow-coupled environments. It establishes a direct connection between biological flight behavior and control theory, and provides insights for the design of energy-efficient autonomous systems operating under environmental uncertainty.

Several directions may further extend this framework. First, extending from point-based sensing to spatial and temporal perception is critical. Incorporating distributed measurements \cite{reddy2016learning} and temporal memory \cite{kim2024wing} may enable the agent to resolve more stochastic flow structures. Second, integrating active propulsion would allow exploration of hybrid flight strategies, such as flap--gliding \cite{kempton2022optimization, shamoun2016flap}, and enable operation in low-energy environments where pure dynamic soaring is insufficient. Third, experimental validation through real-world deployment remains an essential step toward practical applications \cite{reddy2018glider}.

\section{Methods}
\label{sec:methods}

\subsection{Simulation Model}
\label{subsec:model}

The agent is modeled as a 3-degree-of-freedom (3-DOF) point-mass glider, a standard approximation for studying the energy-harvesting trajectories of the wandering albatross (\textit{Diomedea exulans}) \cite{sachs2005minimum, deittert2009engineless, bousquet2017optimal}. 
The glider dynamics are represented by a six-dimensional state vector $\mathbf{s} = [u, \theta, \psi, x, y, z]^T$. The wind vector is defined as $\mathbf{w} = [w(z) \cos \psi_w, w(z) \sin \psi_w, 0]^T$, where $\psi_w$ represents the wind direction. The governing equations are derived as follows:
\begin{align}
    \dot{u} &= -D/m - g \sin \theta - \dot{w} \cos \theta \cos (\psi - \psi_w) \\
    \dot{\theta} &=  \left[ {L \cos \phi}/{m} - g \cos \theta + \dot{w} \sin \theta \cos (\psi - \psi_w) \right] / {u}
    \label{eq:EOM_dot_theta} \\
    \dot{\psi} &= \left[ {L \sin \phi}/{m} + \dot{w} \sin (\psi - \psi_w) \right] / ({u \cos \theta})
    \label{eq:EOM_dot_psi} \\
    \dot{x} &= u \cos \theta \cos \psi + w \cos \psi_w \\
    \dot{y} &= u \cos \theta \sin \psi + w \sin \psi_w \\
    \dot{z} &= u \sin \theta
\end{align}
where $L$ and $D$ are the lift and drag forces, and $m$ is the mass. All numerical values are consistent with Ref.\cite{chen2025optimal}.
The characteristic velocity \(v_c\), length  \(l_c\), and time \(t_c\), can be further defined \cite{bousquet2017optimal}.
The bank angle is constrained to $\phi \in [-80^\circ, 80^\circ]$, and lift coefficient is bounded by $C_L \in [-0.2, 1.5]$, allowing for the high-load, steep-bank turns characteristic of dynamic soaring \cite{sachs2005minimum}.
The term $\dot{w}$ represents the rate of change of the wind speed perceived by the flyer due to its vertical motion through the shear layer:
\begin{equation}
    \dot{w} = \frac{\mathrm{d}w}{\mathrm{d}z} \dot{z} \,.
\end{equation}

The logistic wind profile is set to represent the vertical shear layer:
\begin{equation}
    w(z) = \frac{w_{\mathrm{ref}}}{1 + \exp\left(-\frac{z - z_0}{\delta}\right)} \, ,
\end{equation}
where $w(z)$ is the horizontal wind speed at altitude $z$, $w_{\mathrm{ref}}$ is the reference wind speed above the shear layer, $z_0$ is the inflection point height (representing the center of the shear layer), and $\delta$ characterizes the shear thickness. The corresponding vertical wind gradient, $\sigma(z)$, provides the essential energy source for the agent \cite{chen2025optimal}:
\begin{equation}
    \sigma(z) = \frac{w_{\mathrm{ref}}}{\delta} \frac{\exp\left(-\frac{z - z_0}{\delta}\right)}{\left[1 + \exp\left(-\frac{z - z_0}{\delta}\right)\right]^2} \,.
\end{equation}

To ensure the agent learns a robust and generalizable control policy, the environment parameters are chosen carefully based on a combination of climatological data and aerodynamic scaling laws.

The reference wind speed $w_{\mathrm{ref}}$ is uniformly sampled from $[6,\,20]\,\mathrm{m/s}$. The lower bound ensures feasibility of omnidirectional flight under finite-thickness shear layers, for which realistic thresholds exceed the idealized value of $\sim 3.7\,\mathrm{m/s}$ by $50\%$ \cite{bousquet2017optimal, richardson2022observations}. The upper bound corresponds to the high-wind regime (P90) in the Southern Ocean \cite{derkani2020wind}.

The shear-layer thickness $\delta$ is sampled from $[0.55,\,1.17]\,\mathrm{m}$. The lower bound is derived from geometric constraints of the flyer: requiring the shear layer to be resolvable at the wingspan scale ($6\delta \gtrsim b \approx 3.3\,\mathrm{m}$) yields $\delta \gtrsim 0.55\,\mathrm{m}$ \cite{bousquet2017optimal}. The upper bound maintains the thin-shear regime ($\delta \lesssim 7/6\,\mathrm{m}$) required for efficient energy extraction \cite{bousquet2017optimal, pennycuick2002gust}.

The shear-layer center $z_0$ is coupled to $\delta$ as $z_0 \in [3\delta,\,6\delta]$, ensuring near-zero wind at the surface and consistency with wave-induced flow scaling \cite{buckley2025direct}.

\subsection{Model-free DRL}
\label{subsec:RL_framework}

We formulate the problem as a Markov decision process within a deep reinforcement learning framework (\autoref{fig:problem}D) \cite{adamski2023towards, park2025application}. The agent (glider) learns a policy $\pi(a_t \mid o_t)$ that maps real-time observations $o_t$ to continuous control actions $a_t=(\phi, C_L)$.
The policy is optimized to maximize the discounted return $r=\sum_{t=0}^{N} \gamma^t r_t$ using the Soft Actor–Critic (SAC) algorithm \cite{haarnoja2018soft}. Curriculum learning is employed to stabilize training \cite{bengio2009curriculum}.

\paragraph{Initialization}
\label{subsubsec:initialization}

To balance exploration with solvability in long-horizon soaring tasks, the agent's initial state $\mathbf{s}_0$ and action $\mathbf{a}_0$ are initialized within a physically viable envelope.

\textit{State Initialization ($\mathbf{s}_0$).}
At the beginning of each training episode, the state vector $\mathbf{s}_0$ is initialized with controlled randomization to prevent over-fitting \cite{tobin2017domain} while ensuring feasibility. 
To ensure sufficient initial lift, $u_0$ is sampled from $\max(v_\mathrm{c}, w_{\mathrm{ref}}) \times [0.9, 1.1]$. The flight-path angle $\theta_0$ is sampled near-horizontal within $\pm 10^\circ$. The heading $\psi_0$ is biased towards a crosswind orientation, sampled as $\psi_w + 90^\circ \pm 30^\circ$.
The initial altitude $z_0$ is set relative to the randomized shear layer: $z_0 \in [z_0, z_0 + 2\delta]$, ensuring the agent is initialized within the active region of the wind gradient.

\textit{Action Initialization ($\mathbf{a}_0$).} To prevent the simulation from beginning in an unstable or diverging aerodynamic regime. The lift coefficient is initialized as $C_{L,0} \in [0.5, 1.2]$, representing a moderate-to-high lift state, while the bank angle is sampled from $\phi_0 \in [-5^\circ, 5^\circ]$ to maintain a near-level wings-level attitude. 

\paragraph{Decision frequency}
\label{subsubsec:t_decision}

To ensure stability and biological realism, we decouple the simulation timestep \(t_{\mathrm{sim}}\) from the agent's decision frequency \(t_{\mathrm{decision}}\). The dynamics are integrated with \(t_{\mathrm{sim}} = t_c/100 \approx 14.8\,\mathrm{ms}\) using an explicit Euler scheme \cite{reddy2016learning, flato2024revealing}. The agent policy updates every \(k=4\) steps, yielding a decision interval \(t_{\mathrm{decision}} = k\,t_{\mathrm{sim}} \approx 59.2\,\mathrm{ms}\), which aligns with avian neuro–motor response times (e.g., \(38\)–\(80\,\mathrm{ms}\)) \cite{potts1984chorus, pomeroy1977laboratory, barate2006design}. This prevents exploitation of high-frequency artifacts and encourages robust, high-level soaring strategies.

\paragraph{Observation design}
\label{subsubsec:obs_design}

The base observation space is designed to support simultaneous energy harvesting and goal-directed navigation. It includes (i) relative horizontal displacement $(\Delta x, \Delta y)$ to encode target direction, (ii) altitude $z$ to prevent ground collision, (iii) horizontal wind velocity $(w_x, w_y)$ and vertical wind gradient $\sigma$ to characterize the flow field, and (iv) airspeed components $(u_x, u_y, u_z)$ to represent the aerodynamic state.

\textit{Observation frames.}
We consider both geocentric and egocentric representations \cite{jiao2025sensing}. The geocentric frame $(\cdot)_g$ is Earth-fixed \cite{mouritsen2018long}, whereas the egocentric frame $(\cdot)_e$ is aligned with the horizontal projection of the airspeed vector \cite{yechout2003introduction, mohamed2014fixed, o2022neural}. In this study, frame differences are restricted to the horizontal plane, with a shared vertical axis. The schematics of these frames are shown in \autoref{fig:problem}.

\textit{Coordinate representation.}
Wind observations are expressed either in Cartesian form $(w_{x,e}, w_{y,e})$ or in polar form $(w, \psi_{w,e})$. These representations are mathematically equivalent but differ in their suitability for policy learning.

\textit{Speed representation.}
We compare airspeed- and groundspeed-based observation manifolds. The airspeed formulation $(\mathbf{u}, \mathbf{w})$ provides direct access to aerodynamic variables governing lift, drag, and stall limits, whereas the groundspeed formulation $(\mathbf{v}, \mathbf{w})$ directly encodes navigation progress but requires implicit inference of aerodynamic state.

\paragraph{Reward design}
\label{subsubsec:rwd_design}

The reward structure consists of three components:
\begin{equation}
    r_t = r_{\mathrm{terminal}} + r_{\mathrm{constraint}} + r_{\mathrm{process}} .
\end{equation}
The terminal reward $r_{\mathrm{terminal}}$ enforces mission completion and safety boundaries. A reward of $+20.0$ is granted when the agent enters the $2l_c$ target radius. A crash penalty of $-20.0$ is applied if the altitude falls below the safety threshold ($z < z_{\min}$), and a timeout penalty of $-15.0$ is imposed when the flight duration exceeds $N$.

To ensure biological plausibility, we impose a load-factor constraint \(r_{\mathrm{constraint}} = \xi_l (n-1)\) if \(n>3\), which penalizes excessive aerodynamic load factors $n$ beyond the physiological limits of wandering albatross flight \cite{sachs2005minimum}. The coefficient $\xi_l$ controls the weight of this penalty.

The process reward $r_{\mathrm{process}}$ is designed to guide the agent during flight and is implemented in two alternative forms with different levels of physical abstraction.

The first formulation is \emph{process-based} and directly encodes physically interpretable flight coefficients:
\begin{equation}
    r_{\mathrm{process}} =
    \xi_{\dot{e}}\frac{\dot{e}}{e_{\mathrm{norm}} \sigma_{\mathrm{norm}}}
    + \xi_{v}\,\frac{v_{\mathrm{net}}}{u},
    \label{eq:r_process_A}
\end{equation}
where the energy harvest rate, \(\dot{e}=0.5\,mu^2\sigma(z)\sin(2\theta)\sin(\psi)\) \cite{kempton2022optimization}, \(v_{\mathrm{net}}=u\cos\theta\cos(\psi-\psi_{t,l})+w\cos(\psi_w-\psi_{t,l}) \) (\autoref{fig:problem}B). In \autoref{eq:r_process_A}, the shear normalization factor is defined as $\sigma_{\mathrm{norm}}=u_c/\delta_{\min}$. $e_{\mathrm{norm}}=0.5\,m u_c^2$. The coefficients $\xi_{\dot{e}}$ and $\xi_{v}$ determine the respective weights of the energy-harvesting and directional components.

The second formulation is \emph{state-change-based} and rewards net outcomes rather than prescribing explicit flight coefficients:
\begin{equation}
    r_{\mathrm{process}} =
    \xi_{e} \frac{\Delta e}{e_{\mathrm{norm}}}
    +
    \xi_{d} \frac{\Delta d}{d_{\mathrm{norm}}}.
    \label{eq:r_process_B}
\end{equation}
Here $\Delta e$ denotes the mechanical energy increment and $\Delta d$ represents the distance progress toward the target during one physical decision step $t_{\mathrm{decision}}$. $d_{\mathrm{norm}}=l_c$. The coefficients $\xi_e$ and $\xi_d$ control the relative importance of energy gain and navigational progress.

\paragraph{Curriculum learning}
\label{subsubsec:curriculum_learning}

To enable learning across the full $0^\circ$--$180^\circ$ task space, we employ a curriculum strategy \cite{bengio2009curriculum} that progressively expands the target-direction distribution. Training is initialized over a narrow range ($\psi_t \in [80^\circ,100^\circ]$) and gradually extended to the full interval.
Direct training with a uniform $0^\circ$--$180^\circ$ distribution leads to biased policies that favor intermediate directions ($30^\circ$--$150^\circ$), resulting in poor boundary performance (tailwind and headwind), where success rates fall below $40\%$. To mitigate this, we expand the sampling range to $[-20^\circ,200^\circ]$, converting boundary conditions into interior samples of a wider distribution. This increases data density near the boundaries, improves learning stability, and yields consistent success rates above $95\%$ across the full range.

\paragraph{Algorithm}
\label{para:algorithm}

We employ the SAC algorithm, an off-policy actor–critic method based on the maximum-entropy framework \cite{haarnoja2018soft}. Both actor and critic are implemented as multi-layer perceptrons. We evaluate multiple architectures (see \autoref{tab:NN_test}) and adopt a symmetric $512 \times 512 \times 512$ network as the default configuration.

Angular observations are embedded using trigonometric encoding, $\psi_{w,e} \mapsto [\sin\psi_{w,e}, \cos\psi_{w,e}]^T$, to remove discontinuities at $\pm\pi$ and ensure a smooth representation of periodic variables.

To improve training stability in long-horizon tasks, we employ Leaky ReLU activations ($\alpha=0.01$) \cite{maas2013rectifier}, which maintain non-vanishing gradients in low-activation regimes and preserve sensitivity to rare but critical failure states.

Optimization is performed using Adam with a learning rate of $3\times10^{-5}$. Gradient clipping (maximum norm $0.5$) and weight decay ($10^{-5}$) are applied to stabilize training. The replay buffer size is $10^6$, and a batch size of $4096$ is used to reduce gradient variance. Training runs for up to $2\times10^8$ environment steps.

The equations of motion are integrated in double precision (64-bit), while neural-network computations use single precision (32-bit). Simulations were performed on a high-performance cluster utilizing NVIDIA RTX A4000 GPUs and AMD EPYC CPUs, with an average training time of approximately 0.3--0.6 hours per million environment steps.

\subsection{Statistical Indices}

\paragraph{Success ratio, SR}
\label{para:SR}

Policy performance is evaluated using the Training Success Rate (Training SR) and the Test Success Rate (Test SR).
Training SR is defined as the mean success rate across five independent runs during the steady-state phase ($1.5\times10^8$–$2.0\times10^8$ timesteps), with variance used to quantify training stability.
Test SR evaluates policy robustness. Five checkpoints ($1.6$, $1.7$, $1.8$, $1.9$, and $2.0\times10^8$ timesteps) are selected from the run closest to the ensemble mean. Each checkpoint is evaluated over $1{,}000$ Monte Carlo trials under full stochastic conditions (($\psi_t\in[0^\circ,180^\circ]$, $w_{\mathrm{ref}}\in[6,20]\,\mathrm{m/s}$, $\delta\in[0.55,1.17]\,\mathrm{m}$)).

\paragraph{Transition time}
\label{para:t_star}
The transition time $t^*$ between DS and TG phases is identified based on the spatial localization of energy extraction. In the adopted wind model, energy harvesting is proportional to the local shear magnitude $\sigma$, which peaks near the shear-layer center $z_0$ \cite{kempton2022optimization, chen2025optimal}. 
A trajectory is considered to exit the DS phase when the altitude remains continuously outside the shear region, defined as $z \notin [z_0-2\delta,\, z_0+2\delta]$, for a duration of $3t_c$. This threshold corresponds to regions where the shear magnitude is below approximately $10\%$ of its maximum value.

\paragraph{Two-phase significance ratio, SiR}
\label{para:SiR}

To quantify the prevalence of two-phase behavior, trajectories are sampled under full stochastic conditions. For each successful trajectory, $t^*$ is determined and the trajectory is partitioned into DS and TG phases. A Kolmogorov–Smirnov test is then applied to compare the altitude distributions of the two phases. The proportion of trajectories with statistically significant separation ($p<0.01$) is defined as the two-phase significance ratio, SiR.

\subsection{Generalization Setup}
\label{subsec:gen_setup}

For spatially varying wind fields, wind parameters $\Phi \in \{w_\mathrm{ref}, \delta, \psi_w\}$ are modulated as
\begin{equation}
    \Phi(\mathbf{p}) = \Phi_0 + \mathcal{A}_\Phi M H(\mathbf{p}),
\end{equation}
where $H(\mathbf{p}) = \cos\!\left(\pi x/l\right)\cos\!\left(\pi y/l\right)$, $\mathbf{p}=(x,y)$, and $M \in [0,1]$ controls disturbance intensity. The nominal parameters are $ \Phi_{0,\psi_{w}}=90^\circ$, $\Phi_{0,w_{\mathrm{ref}}}=13\,\mathrm{m/s}$, and $\Phi_{0,\delta} = 0.85\,\mathrm{m}$ with variation amplitudes $\mathcal{A}_{\psi_{w}}=90^\circ$, $\mathcal{A}_{w_{\mathrm{ref}}}=7\,\mathrm{m/s}$, and $\mathcal{A}_\delta=0.35\,\mathrm{m}$. The spatial scale $l$ ranges from $[l_c, d_t]$ ($[21, 512]\,\mathrm{m}$).

For moving-target tasks, the goal follows
\begin{align}
x_t(t) &= d_t + v_{t,m} t \cos \psi_{t,m},\\
y_t(t) &= v_{t,m} t \sin \psi_{t,m},
\end{align}
with velocity $v_{t,m} \in [0,16]\,\mathrm{m/s}$ and heading $\psi_{t,m} \in [-90^\circ, 90^\circ]$.

Gaussian noise is injected at each time step:
\begin{equation}
    \tilde{o}_t = o_t + \xi_t, \quad \xi_t \sim \mathcal{N}(0, \sigma^2 I),
\end{equation}
where $o_t$ is the normalized observation and $\sigma$ controls noise intensity.

\section{Additional information}

\paragraph{Author contributions}
Conceptualization, L.C.; Methodology, L.C.; Investigation, L.C.; Original Draft, L.C.; Review \& Editing, L.C., J.L., Y.Y., and J.H.; Funding Acquisition, Y.X. and H.L.; Resources, Y.X. and H.L.; Supervision, Y.X. and H.L.

\paragraph{Competing interests}
The authors declare no competing financial interests.

\paragraph{Data availability}
Correspondence and requests for materials should be addressed to Yang Xiang (\href{mailto:xiangyang@sjtu.edu.cn}{xiangyang@sjtu.edu.cn}) or Hong Liu (\href{mailto:hongliu@sjtu.edu.cn}{hongliu@sjtu.edu.cn}).

\bibliographystyle{unsrt}
\bibliography{reference}

@article{flato2024revealing,
  title={Revealing principles of autonomous thermal soaring in windy conditions using vulture-inspired deep reinforcement-learning},
  author={Flato, Yoav and Harel, Roi and Tamar, Aviv and Nathan, Ran and Beatus, Tsevi},
  journal={Nature Communications},
  volume={15},
  number={1},
  pages={4942},
  year={2024},
  publisher={Nature Publishing Group UK London}
}

@article{yonehara2016flight,
  title={Flight paths of seabirds soaring over the ocean surface enable measurement of fine-scale wind speed and direction},
  author={Yonehara, Yoshinari and Goto, Yusuke and Yoda, Ken and Watanuki, Yutaka and Young, Lindsay C and Weimerskirch, Henri and Bost, Charles-Andr{\'e} and Sato, Katsufumi},
  journal={Proceedings of the National Academy of Sciences},
  volume={113},
  number={32},
  pages={9039--9044},
  year={2016},
  publisher={National Academy of Sciences}
}

@article{stuber2025miniature,
  title={Miniature multihole airflow sensor for lightweight aircraft over wide speed and angular range},
  author={Stuber, Lukas and Jeger, Simon Luis and Zufferey, Raphael and Floreano, Dario},
  journal={IEEE Robotics and Automation Letters},
  year={2025},
  publisher={IEEE}
}

@inproceedings{bousquet2017dynamic,
  title={Dynamic Soaring in Finite-Thickness Wind Shears: an Asymptotic Solution},
  author={Bousquet, Gabriel D and Triantafyllou, Michael S and Slotine, Jean-Jacques E},
  booktitle={AIAA Guidance, Navigation, and Control Conference},
  pages={1908},
  year={2017}
}

@article{derkani2020wind,
  title={Wind, waves, and surface currents in the Southern Ocean: observations from the Antarctic Circumnavigation Expedition},
  author={Derkani, Marzieh H and Alberello, Alberto and Nelli, Filippo and Bennetts, Luke G and Hessner, Katrin G and MacHutchon, Keith and Reichert, Konny and Aouf, Lotfi and Khan, Salman Saeed and Toffoli, Alessandro},
  journal={Earth System Science Data Discussions},
  volume={2020},
  pages={1--22},
  year={2020},
  publisher={G{\"o}ttingen, Germany}
}

@article{buckley2025direct,
  title={Direct observations of airflow separation over ocean surface waves},
  author={Buckley, Marc P and Horstmann, Jochen and Savelyev, Ivan and Carpenter, Jeff R},
  journal={Nature Communications},
  volume={16},
  number={1},
  pages={5526},
  year={2025},
  publisher={Nature Publishing Group UK London}
}

@article{bousquet2017optimal,
  title={Optimal dynamic soaring consists of successive shallow arcs},
  author={Bousquet, Gabriel D and Triantafyllou, Michael S and Slotine, Jean-Jacques E},
  journal={Journal of The Royal Society Interface},
  volume={14},
  number={135},
  pages={20170496},
  year={2017},
  publisher={The Royal Society}
}

@article{goto2022did,
  title={How did extinct giant birds and pterosaurs fly? A comprehensive modeling approach to evaluate soaring performance},
  author={Goto, Yusuke and Yoda, Ken and Weimerskirch, Henri and Sato, Katsufumi},
  journal={PNAS nexus},
  volume={1},
  number={1},
  pages={pgac023},
  year={2022},
  publisher={Oxford University Press}
}

@article{chen2025optimal,
  title={Optimal dynamic soaring trades off energy harvest and directional flight},
  author={Chen, Lunbing and Yin, Yufei and Xiang, Yang and Qin, Suyang and Liu, Hong},
  journal={iScience},
  volume={28},
  number={6},
  year={2025},
  publisher={Elsevier}
}

@article{weimerskirch2000fast,
  title={Fast and fuel efficient? Optimal use of wind by flying albatrosses},
  author={Weimerskirch, Henri and Guionnet, T and Martin, JSSA and Shaffer, Scott A and Costa, DP},
  journal={Proceedings of the Royal Society of London. Series B: Biological Sciences},
  volume={267},
  number={1455},
  pages={1869--1874},
  year={2000},
  publisher={The Royal Society}
}

@article{zhao2004optimal,
  title={Optimal patterns of glider dynamic soaring},
  author={Zhao, Yiyuan J},
  journal={Optimal control applications and methods},
  volume={25},
  number={2},
  pages={67--89},
  year={2004},
  publisher={Wiley Online Library}
}

@book{anderson2011ebook,
  title={EBOOK: Fundamentals of Aerodynamics (SI units)},
  author={Anderson, John},
  year={2011},
  publisher={McGraw hill}
}

@article{pomeroy1977laboratory,
  title={Laboratory determination of startle reaction time of the starling (Sturnus vulgaris)},
  author={Pomeroy, Harold and Heppner, Frank},
  journal={Animal Behaviour},
  volume={25},
  pages={720--725},
  year={1977},
  publisher={Elsevier}
}

@article{potts1984chorus,
  title={The chorus-line hypothesis of manoeuvre coordination in avian flocks},
  author={Potts, Wayne K},
  journal={Nature},
  volume={309},
  number={5966},
  pages={344--345},
  year={1984},
  publisher={Nature Publishing Group UK London}
}

@article{barate2006design,
  title={Design of a bio-inspired controller for dynamic soaring in a simulated unmanned aerial vehicle},
  author={Barate, Renaud and Doncieux, St{\'e}phane and Meyer, Jean-Arcady},
  journal={Bioinspiration \& biomimetics},
  volume={1},
  number={3},
  pages={76},
  year={2006},
  publisher={IOP Publishing}
}

@inproceedings{park2025application,
  title={Application of Reinforcement Learning for Autonomous Dynamic Soaring},
  author={Park, Sungje and Fanjoy, Adrian and Golubev, Vladimir V},
  booktitle={AIAA SCITECH 2025 Forum},
  pages={2290},
  year={2025}
}

@article{taylor2016soaring,
  title={Soaring energetics and glide performance in a moving atmosphere},
  author={Taylor, Graham K and Reynolds, Kate V and Thomas, Adrian LR},
  journal={Philosophical Transactions of the Royal Society B: Biological Sciences},
  volume={371},
  number={1704},
  pages={20150398},
  year={2016},
  publisher={The Royal Society}
}

@article{harms2025robust,
  title={Robust Optimization-based Autonomous Dynamic Soaring with a Fixed-Wing UAV},
  author={Harms, Marvin and Lim, Jaeyoung and Rohr, David and Rockenbauer, Friedrich and Lawrance, Nicholas and Siegwart, Roland},
  journal={arXiv preprint arXiv:2512.06610},
  year={2025}
}

@article{darveniza2026larval,
  title={Larval zebrafish minimize energy consumption during hunting via adaptive movement selection},
  author={Darveniza, Thomas and Wong, Robert and Zhu, Shuyu I and Pujic, Zac and Sun, Biao and Levendosky, Matthew and Agarwal, Ramesh and McCullough, Michael H and Goodhill, Geoffrey J},
  journal={Proceedings of the National Academy of Sciences},
  volume={123},
  number={7},
  pages={e2513853123},
  year={2026},
  publisher={National Academy of Sciences}
}

@inproceedings{montella2014reinforcement,
  title={Reinforcement learning for autonomous dynamic soaring in shear winds},
  author={Montella, Corey and Spletzer, John R},
  booktitle={2014 IEEE/RSJ International Conference on Intelligent Robots and Systems},
  pages={3423--3428},
  year={2014},
  organization={IEEE}
}

@inproceedings{abozeid2023comprehensive,
  title={A comprehensive assessment to the potential of reinforcement learning in dynamic soaring},
  author={Abozeid, Sara and Pokhrel, Sameer and Eisa, Sameh},
  booktitle={AIAA SCITECH 2023 Forum},
  pages={2236},
  year={2023}
}

@inproceedings{dipaola2023framework,
  title={A Framework for Developing Robust, Autonomous, Power Managed Dynamic Soaring Flight Controllers Using Deep Reinforcement Learning},
  author={DiPaola, Milo F and Barkin, Tyler F},
  booktitle={AIAA AVIATION 2023 Forum},
  pages={4046},
  year={2023}
}

@article{reddy2016learning,
  title={Learning to soar in turbulent environments},
  author={Reddy, Gautam and Celani, Antonio and Sejnowski, Terrence J and Vergassola, Massimo},
  journal={Proceedings of the National Academy of Sciences},
  volume={113},
  number={33},
  pages={E4877--E4884},
  year={2016},
  publisher={National Academy of Sciences}
}

@article{reddy2018glider,
  title={Glider soaring via reinforcement learning in the field},
  author={Reddy, Gautam and Wong-Ng, Jerome and Celani, Antonio and Sejnowski, Terrence J and Vergassola, Massimo},
  journal={Nature},
  volume={562},
  number={7726},
  pages={236--239},
  year={2018},
  publisher={Nature Publishing Group UK London}
}

@article{sachs2012flying,
  title={Flying at no mechanical energy cost: disclosing the secret of wandering albatrosses},
  author={Sachs, Gottfried and Traugott, Johannes and Nesterova, Anna P and Dell'Omo, Giacomo and K{\"u}mmeth, Franz and Heidrich, Wolfgang and Vyssotski, Alexei L and Bonadonna, Francesco},
  year={2012},
  publisher={Public Library of Science San Francisco, USA}
}

@article{sachs2005minimum,
  title={Minimum shear wind strength required for dynamic soaring of albatrosses},
  author={Sachs, Gottfried},
  journal={Ibis},
  volume={147},
  number={1},
  pages={1--10},
  year={2005},
  publisher={Wiley Online Library}
}

@article{kai2019novel,
  title={Novel approach to dynamic soaring modeling and simulation},
  author={Kai, Jean-Marie and Hamel, Tarek and Samson, Claude},
  journal={Journal of Guidance, Control, and Dynamics},
  volume={42},
  number={6},
  pages={1250--1260},
  year={2019},
  publisher={American Institute of Aeronautics and Astronautics}
}

@article{sachs2019kinetic,
  title={Kinetic energy in dynamic soaring—inertial speed and airspeed},
  author={Sachs, Gottfried},
  journal={Journal of Guidance, Control, and Dynamics},
  volume={42},
  number={8},
  pages={1812--1821},
  year={2019},
  publisher={American Institute of Aeronautics and Astronautics}
}

@article{rayleigh1883soaring,
  title={The soaring of birds},
  author={Rayleigh, Lord},
  journal={Nature},
  volume={27},
  number={701},
  pages={534--535},
  year={1883}
}

@book{sutton1998reinforcement,
  title={Reinforcement learning: An introduction},
  author={Sutton, Richard S and Barto, Andrew G and others},
  volume={1},
  number={1},
  year={1998},
  publisher={MIT press Cambridge}
}

@inproceedings{maas2013rectifier,
  title={Rectifier nonlinearities improve neural network acoustic models},
  author={Maas, Andrew L and Hannun, Awni Y and Ng, Andrew Y and others},
  booktitle={Proc. icml},
  volume={30},
  number={1},
  pages={3},
  year={2013},
  organization={Atlanta, GA}
}

@inproceedings{bengio2009curriculum,
  title={Curriculum learning},
  author={Bengio, Yoshua and Louradour, J{\'e}r{\^o}me and Collobert, Ronan and Weston, Jason},
  booktitle={Proceedings of the 26th annual international conference on machine learning},
  pages={41--48},
  year={2009}
}

@inproceedings{tobin2017domain,
  title={Domain randomization for transferring deep neural networks from simulation to the real world},
  author={Tobin, Josh and Fong, Rachel and Ray, Alex and Schneider, Jonas and Zaremba, Wojciech and Abbeel, Pieter},
  booktitle={2017 IEEE/RSJ international conference on intelligent robots and systems (IROS)},
  pages={23--30},
  year={2017},
  organization={IEEE}
}

@article{richardson2018flight,
  title={Flight speed and performance of the wandering albatross with respect to wind},
  author={Richardson, Philip L and Wakefield, Ewan D and Phillips, Richard A},
  journal={Movement ecology},
  volume={6},
  number={1},
  pages={3},
  year={2018},
  publisher={Springer}
}

@article{deittert2009engineless,
  title={Engineless unmanned aerial vehicle propulsion by dynamic soaring},
  author={Deittert, Markus and Richards, Arthur and Toomer, Chris A and Pipe, Anthony},
  journal={Journal of guidance, control, and dynamics},
  volume={32},
  number={5},
  pages={1446--1457},
  year={2009}
}

@article{verma2018efficient,
  title={Efficient collective swimming by harnessing vortices through deep reinforcement learning},
  author={Verma, Siddhartha and Novati, Guido and Koumoutsakos, Petros},
  journal={Proceedings of the National Academy of Sciences},
  volume={115},
  number={23},
  pages={5849--5854},
  year={2018},
  publisher={National Academy of Sciences}
}

@article{pennycuick2002gust,
  title={Gust soaring as a basis for the flight of petrels and albatrosses (Procellariiformes)},
  author={Pennycuick, Colin J},
  journal={Avian Science},
  volume={2},
  pages={1--12},
  year={2002},
  publisher={European Ornithologists Union}
}

@inproceedings{bird2014closing,
  title={Closing the loop in dynamic soaring},
  author={Bird, John J and Langelaan, Jack W and Montella, Corey and Spletzer, John and Grenestedt, Joachim L},
  booktitle={AIAA Guidance, Navigation, and Control Conference},
  pages={0263},
  year={2014}
}

@article{jones2022physics,
  title={Physics and modeling of large flow disturbances: discrete gust encounters for modern air vehicles},
  author={Jones, Anya R and Cetiner, Oksan and Smith, Marilyn J},
  journal={Annual Review of Fluid Mechanics},
  volume={54},
  number={1},
  pages={469--493},
  year={2022},
  publisher={Annual Reviews}
}

@inproceedings{bronz2021flight,
  title={Flight testing of dynamic soaring Part-2: Open-field inclined circle trajectory},
  author={Bronz, Murat and Gavrilovic, Nikola and Drouin, Antoine and Hattenberger, Gautier and Moschetta, Jean-Marc},
  booktitle={AIAA Aviation 2021 Forum},
  pages={2803},
  year={2021}
}

@inproceedings{langelaan2012wind,
  title={Wind field estimation for autonomous dynamic soaring},
  author={Langelaan, Jack W and Spletzer, John and Montella, Corey and Grenestedt, Joachim},
  booktitle={2012 IEEE International conference on robotics and automation},
  pages={16--22},
  year={2012},
  organization={IEEE}
}

@article{mouritsen2018long,
  title={Long-distance navigation and magnetoreception in migratory animals},
  author={Mouritsen, Henrik},
  journal={Nature},
  volume={558},
  number={7708},
  pages={50--59},
  year={2018},
  publisher={Nature Publishing Group UK London}
}

@book{yechout2003introduction,
  title={Introduction to aircraft flight mechanics: performance, static stability, dynamic stability, and classical feedback control},
  author={Yechout, Thomas R},
  year={2003},
  publisher={Aiaa}
}

@article{mohamed2014fixed,
  title={Fixed-wing MAV attitude stability in atmospheric turbulence—Part 2: Investigating biologically-inspired sensors},
  author={Mohamed, Abdulghani and Watkins, Simon and Clothier, Reece and Abdulrahim, Mujahid and Massey, Kevin and Sabatini, Roberto},
  journal={Progress in Aerospace Sciences},
  volume={71},
  pages={1--13},
  year={2014},
  publisher={Elsevier}
}

@article{o2022neural,
  title={Neural-fly enables rapid learning for agile flight in strong winds},
  author={O’Connell, Michael and Shi, Guanya and Shi, Xichen and Azizzadenesheli, Kamyar and Anandkumar, Anima and Yue, Yisong and Chung, Soon-Jo},
  journal={Science Robotics},
  volume={7},
  number={66},
  pages={eabm6597},
  year={2022},
  publisher={American Association for the Advancement of Science}
}

@inproceedings{haarnoja2018soft,
  title={Soft actor-critic: Off-policy maximum entropy deep reinforcement learning with a stochastic actor},
  author={Haarnoja, Tuomas and Zhou, Aurick and Abbeel, Pieter and Levine, Sergey},
  booktitle={International conference on machine learning},
  pages={1861--1870},
  year={2018},
  organization={Pmlr}
}

@inproceedings{adamski2023towards,
  title={Towards Development of a Dynamic Soaring Capable UAV using Reinforcement Learning},
  author={Adamski, Jacob R and Golubev, Vladimir V and Gudmundsson, Snorri and Kuznetsov, Fedor},
  booktitle={AIAA AVIATION 2023 Forum},
  pages={4455},
  year={2023}
}

@article{richardson2022observations,
  title={Observations and models of across-wind flight speed of the wandering albatross},
  author={Richardson, Philip L and Wakefield, Ewan D},
  journal={Royal Society Open Science},
  volume={9},
  number={11},
  pages={211364},
  year={2022},
  publisher={The Royal Society}
}

@misc{lillicrap2020continuous,
  title={Continuous control with deep reinforcement learning},
  author={Lillicrap, Timothy Paul and Hunt, Jonathan James and Pritzel, Alexander and Heess, Nicolas Manfred Otto and Erez, Tom and Tassa, Yuval and Silver, David and Wierstra, Daniel Pieter},
  year={2020},
  month=sep # "~15",
  publisher={Google Patents},
  note={US Patent 10,776,692}
}

@article{akhtar2026dynamic,
  title={Dynamic soaring in UAVs: a deep reinforcement learning approach},
  author={Akhtar, Mishma and Maqsood, Adnan and Mir, Imran and Gungordu, Baris},
  journal={The Aeronautical Journal},
  pages={1--29},
  year={2026},
  publisher={Cambridge University Press}
}

@article{todorov2002optimal,
  title={Optimal feedback control as a theory of motor coordination},
  author={Todorov, Emanuel and Jordan, Michael I},
  journal={Nature neuroscience},
  volume={5},
  number={11},
  pages={1226--1235},
  year={2002},
  publisher={Nature Publishing Group US New York}
}

@inproceedings{peng2018sim,
  title={Sim-to-real transfer of robotic control with dynamics randomization},
  author={Peng, Xue Bin and Andrychowicz, Marcin and Zaremba, Wojciech and Abbeel, Pieter},
  booktitle={2018 IEEE international conference on robotics and automation (ICRA)},
  pages={3803--3810},
  year={2018},
  organization={IEEE}
}

@article{langelaan2009enabling,
  title={Enabling new missions for robotic aircraft},
  author={Langelaan, Jack W and Roy, Nicholas},
  journal={Science},
  volume={326},
  number={5960},
  pages={1642--1644},
  year={2009},
  publisher={American Association for the Advancement of Science}
}

@article{uesaka2023wandering,
  title={Wandering albatrosses exert high take-off effort only when both wind and waves are gentle},
  author={Uesaka, Leo and Goto, Yusuke and Naruoka, Masaru and Weimerskirch, Henri and Sato, Katsufumi and Sakamoto, Kentaro Q},
  journal={Elife},
  volume={12},
  pages={RP87016},
  year={2023},
  publisher={eLife Sciences Publications, Ltd}
}

@article{goto2024albatrosses,
  title={Albatrosses employ orientation and routing strategies similar to yacht racers},
  author={Goto, Yusuke and Weimerskirch, Henri and Fukaya, Keiichi and Yoda, Ken and Naruoka, Masaru and Sato, Katsufumi},
  journal={Proceedings of the National Academy of Sciences},
  volume={121},
  number={23},
  pages={e2312851121},
  year={2024},
  publisher={National Academy of Sciences}
}

@article{kempton2022optimization,
  title={Optimization of dynamic soaring in a flap-gliding seabird affects its large-scale distribution at sea},
  author={Kempton, James A and Wynn, Joe and Bond, Sarah and Evry, James and Fayet, Annette L and Gillies, Natasha and Guilford, Tim and Kavelaars, Marwa and Juarez-Martinez, Ignacio and Padget, Oliver and others},
  journal={Science advances},
  volume={8},
  number={22},
  pages={eabo0200},
  year={2022},
  publisher={American Association for the Advancement of Science}
}

@article{shamoun2016flap,
  title={Flap or soar? How a flight generalist responds to its aerial environment},
  author={Shamoun-Baranes, Judy and Bouten, Willem and Van Loon, E Emiel and Meijer, Christiaan and Camphuysen, CJ},
  journal={Philosophical Transactions of the Royal Society B: Biological Sciences},
  volume={371},
  number={1704},
  year={2016},
  publisher={The Royal Society}
}

@article{sachs2013experimental,
  title={Experimental verification of dynamic soaring in albatrosses},
  author={Sachs, G and Traugott, J and Nesterova, AP and Bonadonna, F},
  journal={Journal of Experimental Biology},
  volume={216},
  number={22},
  pages={4222--4232},
  year={2013},
  publisher={Company of Biologists}
}

@book{stull2012introduction,
  title={An introduction to boundary layer meteorology},
  author={Stull, Roland B},
  year={2012},
  publisher={Springer Science \& Business Media}
}

@article{goto2017asymmetry,
  title={Asymmetry hidden in birds’ tracks reveals wind, heading, and orientation ability over the ocean},
  author={Goto, Yusuke and Yoda, Ken and Sato, Katsufumi},
  journal={Science advances},
  volume={3},
  number={9},
  pages={e1700097},
  year={2017},
  publisher={American Association for the Advancement of Science}
}

@article{shoval2012evolutionary,
  title={Evolutionary trade-offs, Pareto optimality, and the geometry of phenotype space},
  author={Shoval, Oren and Sheftel, Hila and Shinar, Guy and Hart, Yuval and Ramote, Omer and Mayo, Avi and Dekel, Erez and Kavanagh, Kathryn and Alon, Uri},
  journal={Science},
  volume={336},
  number={6085},
  pages={1157--1160},
  year={2012},
  publisher={American Association for the Advancement of Science}
}

@article{weimerskirch2002gps,
  title={GPS tracking of foraging albatrosses},
  author={Weimerskirch, Henri and Bonadonna, Francesco and Bailleul, Fr{\'e}d{\'e}ric and Mabille, G{\'e}raldine and Dell'Omo, Giacomo and Lipp, Hans-Peter},
  journal={Science},
  volume={295},
  number={5558},
  pages={1259--1259},
  year={2002},
  publisher={American Association for the Advancement of Science}
}

@article{mohamed2022opportunistic,
  title={Opportunistic soaring by birds suggests new opportunities for atmospheric energy harvesting by flying robots},
  author={Mohamed, Abdulghani and Taylor, Graham K and Watkins, Simon and Windsor, Shane P},
  journal={Journal of the Royal Society Interface},
  volume={19},
  number={196},
  pages={20220671},
  year={2022},
  publisher={The Royal Society}
}

@article{hong2023dynamic,
  title={Dynamic soaring under different atmospheric stability conditions},
  author={Hong, Haichao and Liu, Luoqin and Holzapfel, Florian and Sachs, Gottfried},
  journal={Journal of Guidance, Control, and Dynamics},
  volume={46},
  number={5},
  pages={970--977},
  year={2023},
  publisher={American Institute of Aeronautics and Astronautics}
}

@article{jiao2025sensing,
  title={Sensing flow gradients is necessary for learning autonomous underwater navigation},
  author={Jiao, Yusheng and Hang, Haotian and Merel, Josh and Kanso, Eva},
  journal={Nature Communications},
  volume={16},
  number={1},
  pages={3044},
  year={2025},
  publisher={Nature Publishing Group UK London}
}

@article{kim2024wing,
  title={Wing-strain-based flight control of flapping-wing drones through reinforcement learning},
  author={Kim, Taewi and Hong, Insic and Im, Sunghoon and Rho, Seungeun and Kim, Minho and Roh, Yeonwook and Kim, Changhwan and Park, Jieun and Lim, Daseul and Lee, Doohoe and others},
  journal={Nature Machine Intelligence},
  volume={6},
  number={9},
  pages={992--1005},
  year={2024},
  publisher={Nature Publishing Group UK London}
}

@article{gunnarson2021learning,
  title={Learning efficient navigation in vortical flow fields},
  author={Gunnarson, Peter and Mandralis, Ioannis and Novati, Guido and Koumoutsakos, Petros and Dabiri, John O},
  journal={Nature communications},
  volume={12},
  number={1},
  pages={7143},
  year={2021},
  publisher={Nature Publishing Group UK London}
}

@article{notter2023hierarchical,
  title={Hierarchical reinforcement learning approach for autonomous cross-country soaring},
  author={Notter, Stefan and Schimpf, Fabian and M{\"u}ller, Gregor and Fichter, Walter},
  journal={Journal of Guidance, Control, and Dynamics},
  volume={46},
  number={1},
  pages={114--126},
  year={2023},
  publisher={American Institute of Aeronautics and Astronautics}
}

\clearpage

\section*{Supplementary Material}
\setcounter{section}{0}
\setcounter{figure}{0}
\setcounter{table}{0}
\setcounter{equation}{0}
\renewcommand{\thesection}{S\arabic{section}}
\renewcommand{\thefigure}{S\arabic{figure}}
\renewcommand{\thetable}{S\arabic{table}}
\renewcommand{\theequation}{S\arabic{equation}}


\begin{table}[htbp]
    \centering
    \begin{tabular}{ccccc}
    \toprule
    No. & \(\textit{NN}_\text{Actor}\) & \(\textit{NN}_\text{Critic}\) & Training SR & Test SR\\
    \midrule
    1 & \(\left[512,512,512\right]\) & \(\left[512,512,512\right]\) & \(95.5\% \pm 0.7\%\) & \(97.3\% \pm 0.8\%\) \\
    2 & \(\left[512,512\right]\) & \(\left[512,512\right]\) & \( 82.6\% \pm  6.4\%\) & \(82.6\% \pm 2.5\%\)\\
    3 & \(\left[256,256,256\right]\) & \(\left[256,256,256\right]\) & \( 68.2\% \pm  10.4\%\) & \(62.4\% \pm 4.9\%\)\\
    4 & \(\left[512,512,512\right]\) & \(\left[1024,1024,1024\right]\) & \( 91.8\% \pm  3.3\%\) & \(95.8\% \pm 2.7\%\)\\
    \bottomrule
    \end{tabular}
    \caption{
    \textbf{Training and test success rates for different neural network architectures.} All configurations use the same observation space (Obs.E1, \autoref{tab:obs_ablation}) and reward formulation (Rwd.1, \autoref{tab:reward_ablation}). Each architecture is defined by the layer widths of the actor and critic networks.
    }
    \label{tab:NN_test}
\end{table}

\begin{figure}[htbp]
    \centering
    \includegraphics{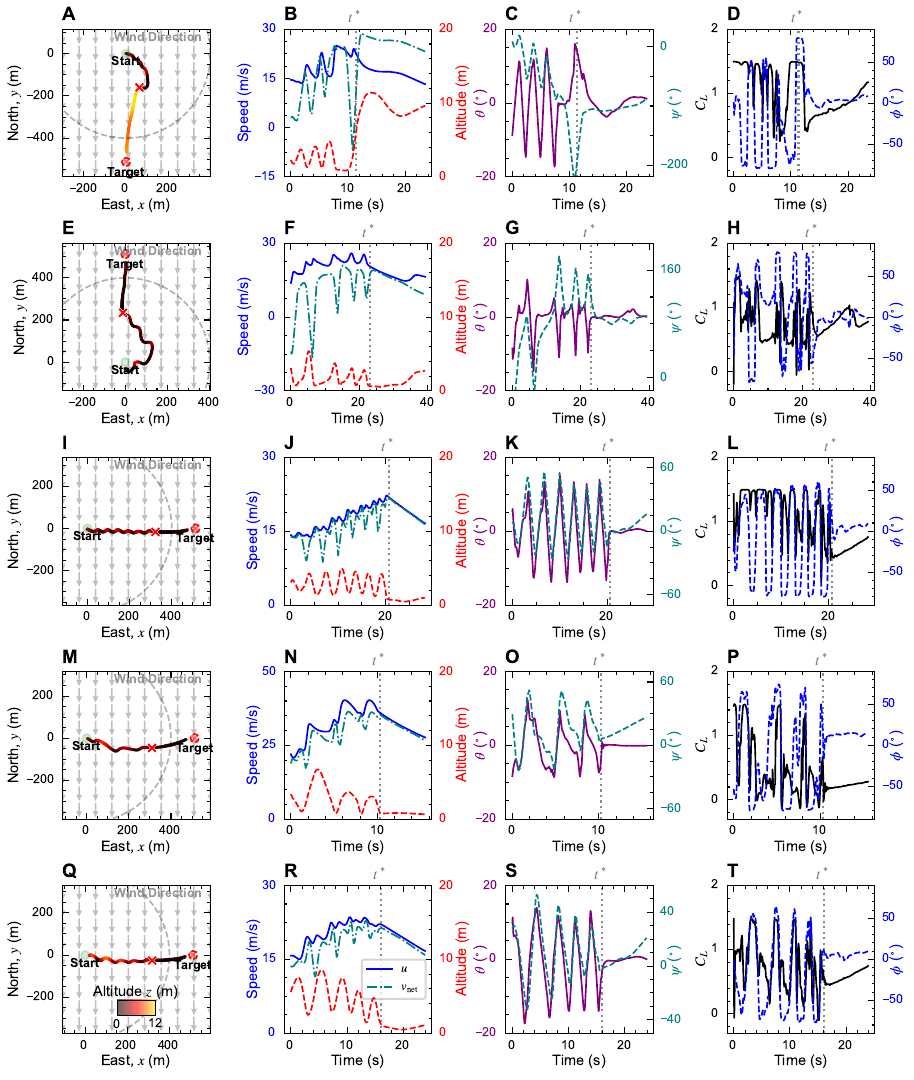}
    \caption{\textbf{Flight trajectories and state variables under representative wind conditions} (corresponding to \autoref{fig:problem}C, \autoref{fig:policy_macro}A–D). 
    The figure presents agent performance across five scenarios: (row 1) tailwind target ($\psi_t=0^\circ,\, w_\mathrm{ref}=10~\mathrm{m/s},\, \delta=0.55~\mathrm{m}$); (row 2) headwind target ($\psi_t=180^\circ,\, w_\mathrm{ref}=10~\mathrm{m/s},\, \delta=0.55~\mathrm{m}$); (row 3) low wind speed ($\psi_t=90^\circ,\, w_\mathrm{ref}=6~\mathrm{m/s},\, \delta=0.55~\mathrm{m}$); (row 4) high wind speed ($\psi_t=90^\circ,\, w_\mathrm{ref}=20~\mathrm{m/s},\, \delta=0.55~\mathrm{m}$); and (row 5) thick shear layer ($\psi_t=90^\circ,\, w_\mathrm{ref}=10~\mathrm{m/s},\, \delta=1.33~\mathrm{m}$). 
    Column 1 shows top-down trajectories in the $x$–$y$ plane, color-coded by altitude $z$. 
    Column 2 presents kinematic variables. 
    Column 3 shows flight angles. 
    Column 4 displays control inputs.}
    \label{fig:S_fig02_1}
\end{figure}

\begin{figure}[htbp]
    \centering
    \includegraphics{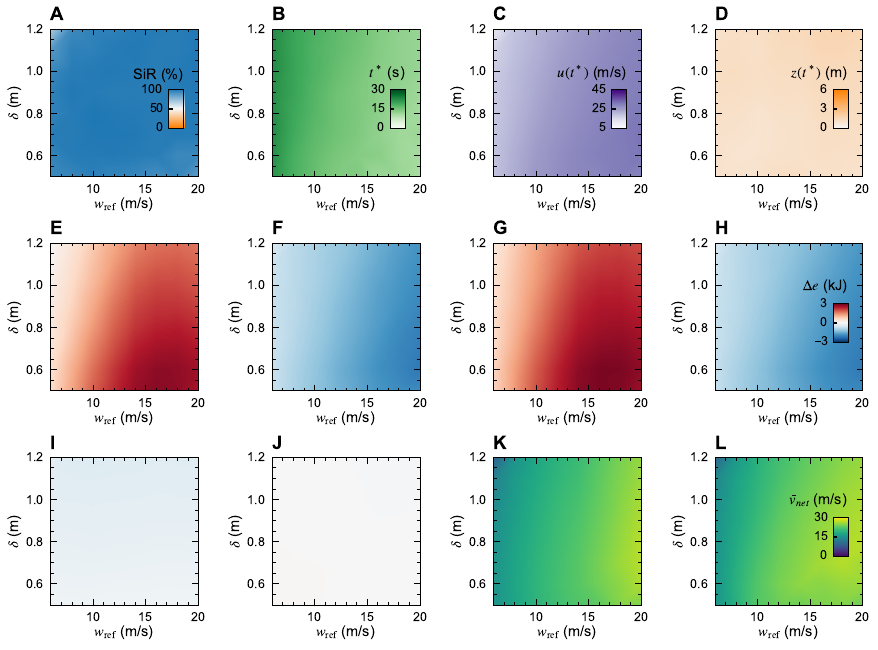}
    \caption{\textbf{Phase-wise performance metrics across wind-shear conditions} ($w_\mathrm{ref}$ and $\delta$; corresponding to \autoref{fig:policy_macro}E–P). 
    \textbf{(A)} Two-phase significance ratio (SiR); \textbf{(B)} transition time $t^*$; \textbf{(C)} airspeed at transition $u(t^*)$; \textbf{(D)} altitude at transition $z(t^*)$. 
    \textbf{(E–L)} Energy and navigation metrics for the two phases. \textbf{(E, F)} net change in total energy $\Delta e$; \textbf{(G, H)} net change in kinetic energy $\Delta e_k$; \textbf{(I, J)} net change in potential energy $\Delta e_p$; \textbf{(K, L)} mean directional velocity $\bar{v}_{\mathrm{net}}$. Panels \textbf{(E, G, I, K)} correspond to the DS phase, and \textbf{(F, H, J, L)} to the TG phase.}
    \label{fig:S_fig02_2}
\end{figure}

\begin{figure}[htbp]
    \centering
    \includegraphics{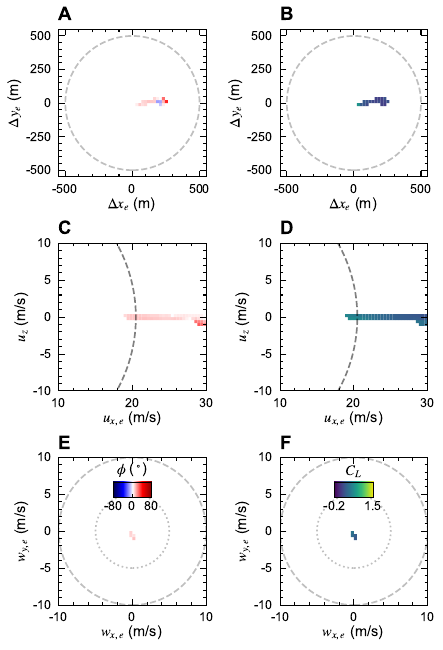}
    \caption{\textbf{Policy distribution during the targeted gliding (TG) phase} (corresponding to \autoref{fig:policy_micro}). 
    The left column (\textbf{A}, \textbf{C}, \textbf{E}) shows the bank angle $\phi$, while the right column (\textbf{B}, \textbf{D}, \textbf{F}) shows the lift coefficient $C_L$.}
    \label{fig:S_fig03_1}
\end{figure}

\begin{figure}[htbp]
    \centering
    \includegraphics{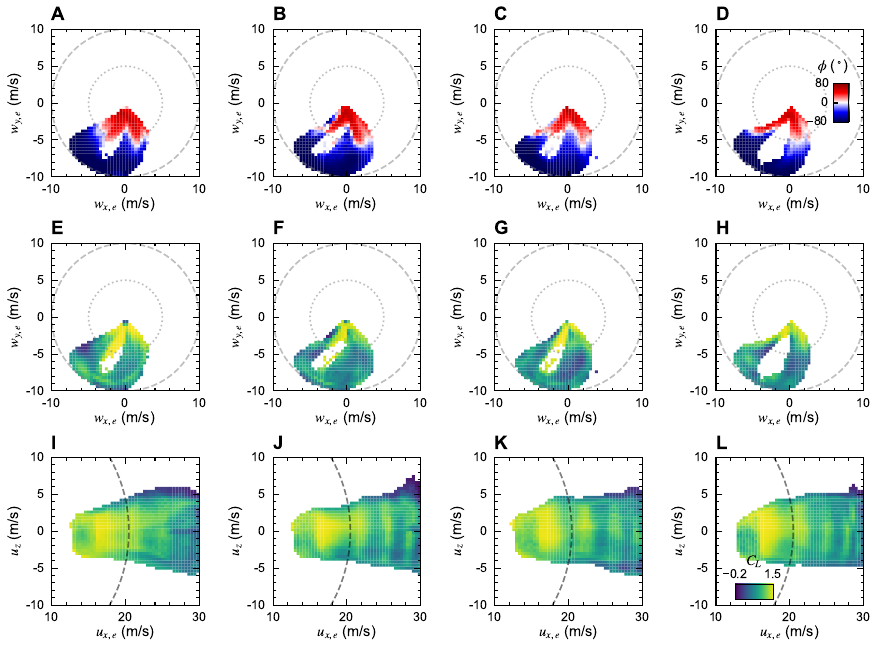}
    \caption{\textbf{Evolution of learned state–action mappings across training.}
    Heatmaps illustrate the policy-induced state–action distributions in multiple kinematic domains at four training checkpoints ($1.6$, $1.7$, $1.8$, and $1.9\times10^8$ timesteps), arranged from left to right.}
    \label{fig:S_fig03_2}
\end{figure}

\begin{figure}[htbp]
    \centering
    \includegraphics{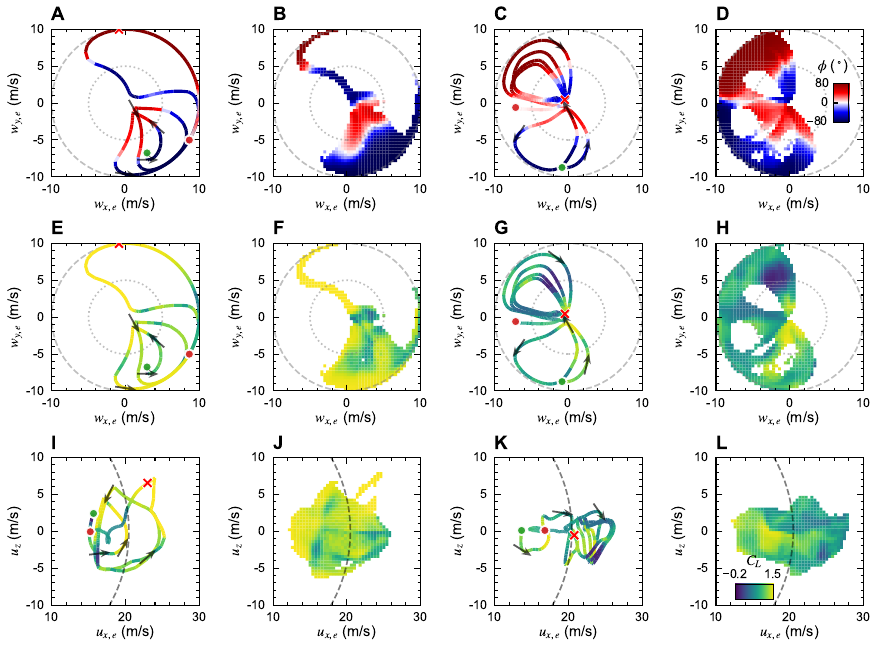}
    \caption{\textbf{Trajectory tracking and policy heatmaps under tailwind and headwind conditions.} 
    Results are shown for the environmental setting $w_\mathrm{ref}=10~\mathrm{m/s}$ and $\delta=0.55~\mathrm{m}$. 
    The left two columns \textbf{(A, B, E, F, I, J)} correspond to a target direction $\psi_t=0^\circ$, while the right two columns \textbf{(C, D, G, H, K, L)} correspond to $\psi_t=180^\circ$. 
    Columns 1 and 3 show representative successful trajectories, whereas columns 2 and 4 present aggregated state–action visitation heatmaps over successful episodes.}
    \label{fig:S_fig03_3}
\end{figure}

\begin{figure}
    \centering
    \includegraphics{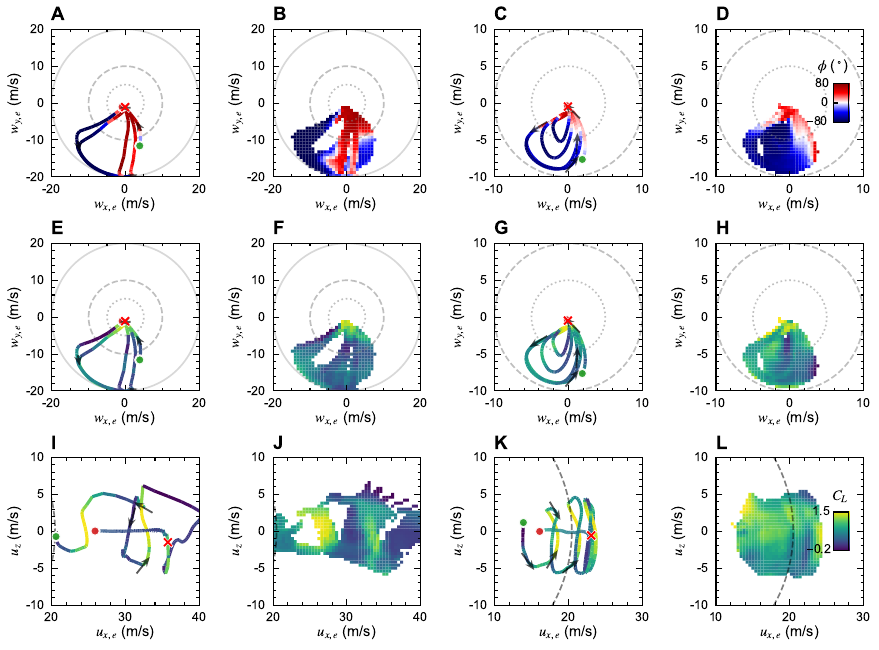}
    \caption{\textbf{Trajectory tracking and policy heatmaps under high-wind and thick-shear conditions.} 
    Results are shown for a fixed target direction $\psi_t=90^\circ$. 
    The left two columns \textbf{(A, B, E, F, I, J)} correspond to $w_\mathrm{ref}=20.0~\mathrm{m/s}$ with $\delta=0.55~\mathrm{m}$, while the right two columns \textbf{(C, D, G, H, K, L)} correspond to $w_\mathrm{ref}=10.0~\mathrm{m/s}$ with $\delta=1.33~\mathrm{m}$. 
    Columns 1 and 3 show representative successful trajectories, whereas columns 2 and 4 present aggregated state–action visitation heatmaps over successful episodes.}
    \label{fig:S_fig03_4}
\end{figure}

\begin{figure}[htbp]
    \centering
    \includegraphics{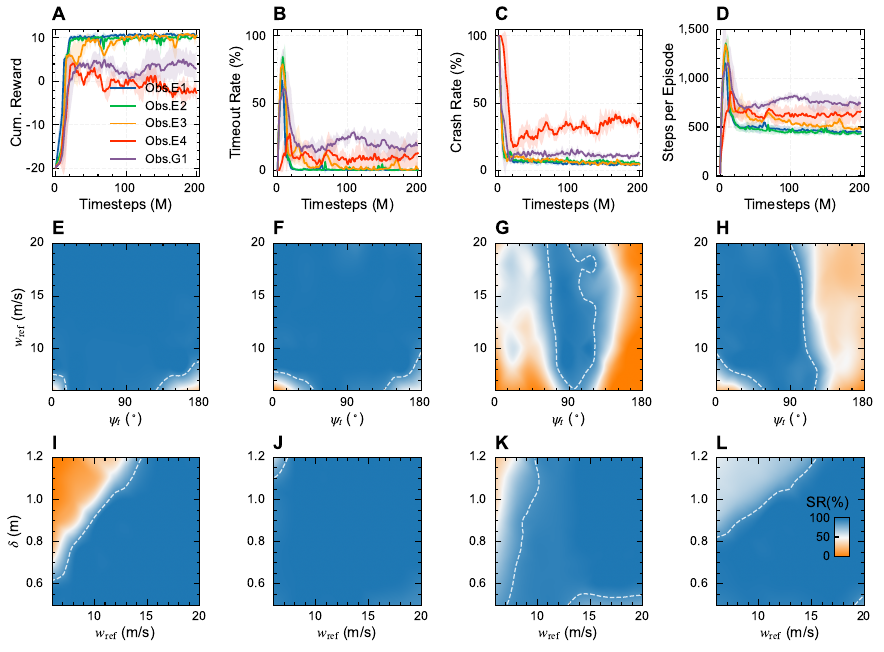}
    \caption{
    \textbf{Training dynamics and observation-space-dependent performance across task and environmental conditions.}
    \textbf{(A–D)} Training curves over timesteps (corresponding to \autoref{fig:problem}G), including cumulative reward (A), timeout rate (B), crash rate (C), and steps per episode (D). Curves correspond to different observation-space designs (\autoref{tab:obs_ablation}); solid lines denote means and shaded regions indicate standard deviations.
    \textbf{(E–L)} Success-rate distributions evaluated within the training domain (corresponding to \autoref{fig:problem}I, J). Columns correspond to Obs.E2 (E,I), Obs.E3 (F,J), Obs.E4 (G,K), and Obs.G1 (H,L). Colormaps represent success rate (SR, \%), with dashed contours indicating regions where SR exceeds 90\%.
    }
    \label{fig:S_fig01_1}
\end{figure}

\begin{figure}[htbp]
    \centering
    \includegraphics{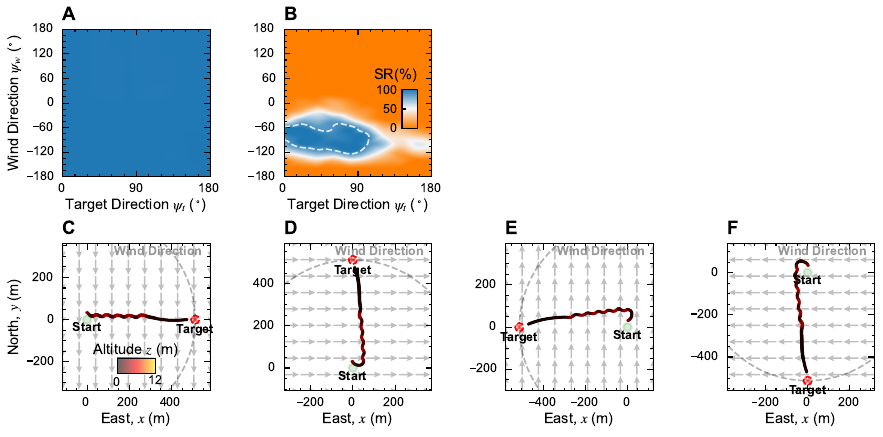}
    \caption{\textbf{Success-rate distributions and trajectory patterns under varying wind directions.}
    \textbf{(A, B)} Contour maps of navigation success rate as a function of target direction ($\psi_t$) and wind direction ($\psi_w$) for egocentric (relative) observations (Obs.E1) (A) and geocentric (absolute) observations (Obs.G1) (B). Dashed contours denote the 90\% success-rate boundary.
    \textbf{(C–F)} Representative trajectories projected onto the horizontal (East–North) plane under different wind directions with $\psi_w = -90^\circ$ (\textbf{C}), $0^\circ$ (\textbf{D}), $90^\circ$ (\textbf{E}), and $180^\circ$ (\textbf{F}), all shown for Obs.E1.
    }
    \label{fig:S_fig01_2}
\end{figure}

\begin{figure}[htbp]
    \centering
    \includegraphics{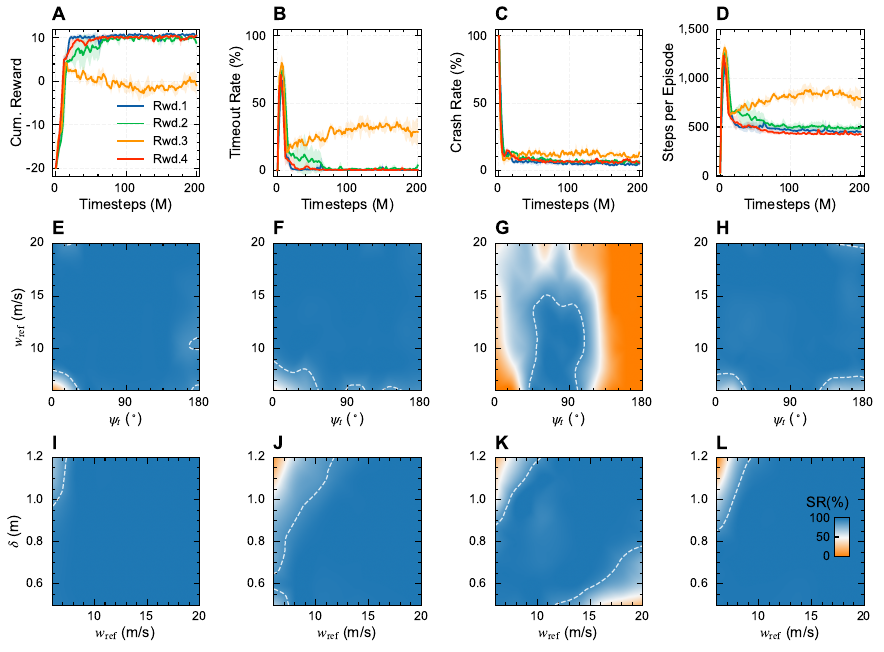}
    \caption{\textbf{Training dynamics and reward-dependent performance across task and environmental conditions.}
    \textbf{(A–D)} Training curves over timesteps (corresponding to \autoref{fig:problem}H), including cumulative reward (A), timeout rate (B), crash rate (C), and steps per episode (D). Curves correspond to different reward designs (\autoref{tab:reward_ablation}); solid lines denote means and shaded regions indicate standard deviations.
    \textbf{(E–L)} Success-rate distributions evaluated within the training domain. Columns correspond to Rwd.1 (E,I), Rwd.2 (F,J), Rwd.3 (G,K), and Rwd.4 (H,L). Colormaps represent success rate (SR, \%), with dashed contours indicating regions where SR exceeds 90\%.
    }
    \label{fig:S_fig01_4}
\end{figure}

\begin{figure}[htbp]
    \centering
    \includegraphics{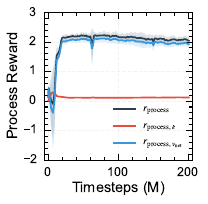}
    \caption{\textbf{Evolution and decomposition of the process reward during training.} 
    Results are shown for Obs.E1 with reward formulation Rwd.1 (\autoref{tab:reward_ablation}). A detailed definition of the process reward and its components is provided in \autoref{subsubsec:rwd_design}.}
    \label{fig:S_fig01_5}
\end{figure}

\begin{figure}[htbp]
    \centering
    \includegraphics{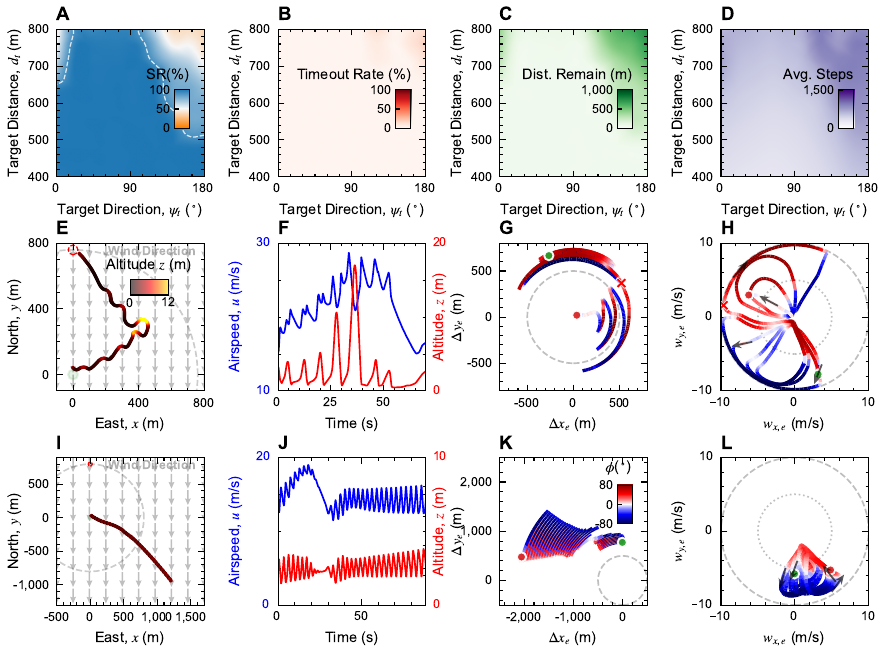}
    \caption{\textbf{Mission feasibility boundaries and limiting behaviors in static target-navigation tasks.} 
    \textbf{(A–D)} Performance landscapes over target direction and distance (corresponding to \autoref{fig:generalization}H), including success rate \textbf{(A)}, timeout rate \textbf{(B)}, terminal distance to the target at episode end \textbf{(C)}, and average episode length \textbf{(D)}. 
    \textbf{(E–H)} Flight analysis for a successful boundary case ($\psi_t=180^\circ,\, d_t=760~\mathrm{m}$). 
    \textbf{(I–L)} Analysis of a timeout case beyond the feasibility limit ($\psi_t=180^\circ,\, d_t=800~\mathrm{m}$). In (I–L), although sustained flight is maintained through repeated energy-harvesting cycles, the agent fails to progress toward the target due to limited exposure to such relative configurations during training (K), resulting in a quasi-stationary loitering pattern until termination.}
    \label{fig:S_fig04_1}
\end{figure}

\begin{figure}[htbp]
    \centering
    \includegraphics{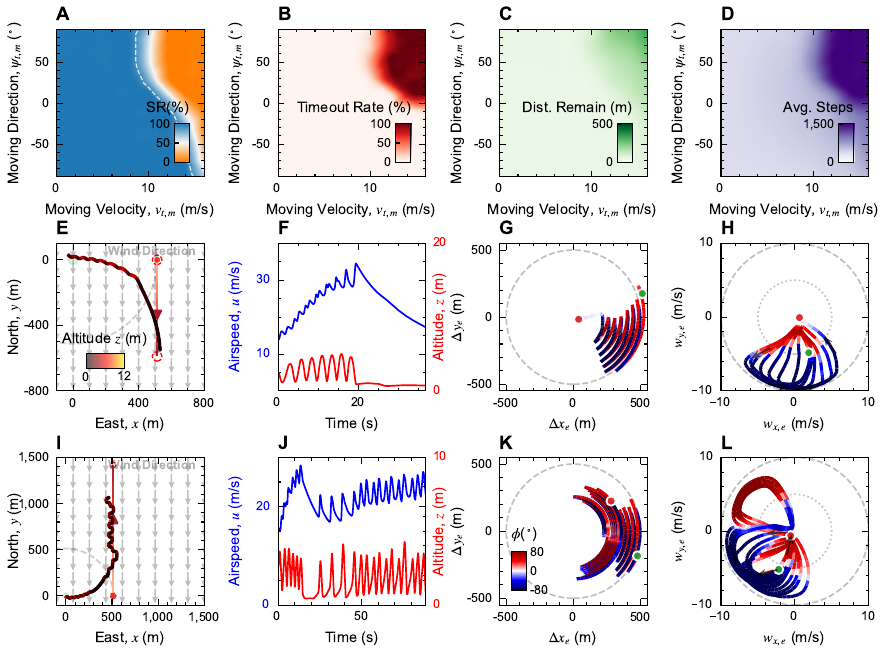}
    \caption{\textbf{Performances of dynamic target interception across velocity and heading regimes.} 
    \textbf{(A–D)} Performance landscapes over target velocity and heading (corresponding to \autoref{fig:generalization}I), including success rate \textbf{(A)}, timeout rate \textbf{(B)}, terminal distance to the target at episode end \textbf{(C)}, and average episode length \textbf{(D)}. 
    \textbf{(E–H)} Successful interception of a fast downwind-moving target ($v_{t,m}=16~\mathrm{m/s}$, $\psi_{t,m}=-90^\circ$).
    \textbf{(I–L)} Interception failure for an identical target moving upwind ($v_{t,m}=16~\mathrm{m/s}$, $\psi_{t,m}=90^\circ$). Although the agent is unable to close the distance, resulting in a timeout-driven chase, it continuously pursues the target.
    In \textbf{(E,I)}, the red arrow indicates the direction of target motion, with endpoints marking the initial and final positions.}
    \label{fig:S_fig04_2}   
\end{figure}

\begin{figure}[htbp]
    \centering
    \includegraphics{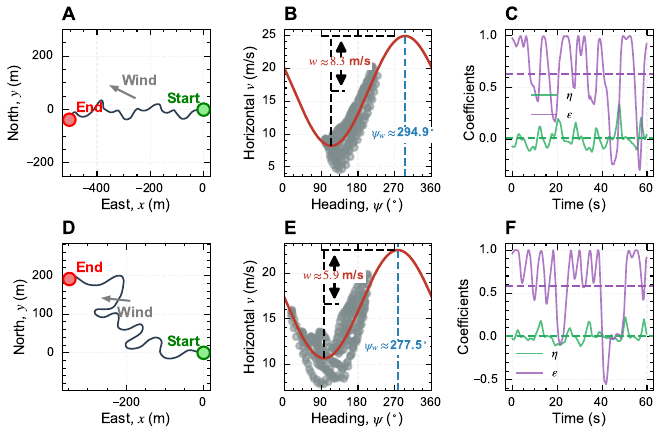}
    \caption{\textbf{Flight trajectories, wind estimation, and energy/direction-related coefficients from empirical seabird data.} 
    Representative samples are from the \textit{BiP\_WA\_CROZET2019\_F0957\_N18035\_G} dataset \cite{uesaka2023wandering}. 
    \textbf{(A, D)} Spatial trajectories of the flights. 
    \textbf{(B, E)} Relationship between heading angle ($\psi$) and horizontal ground speed ($v$). Gray markers denote empirical data, while the solid red curve shows the fitted trigonometric model used to estimate wind speed ($w$) and wind direction ($\psi_w$, indicated by vertical dashed blue lines), following the method of \cite{yonehara2016flight}. 
    \textbf{(C, F)} Temporal evolution of the energy-harvesting rate ($\eta$, green) and the directional efficiency ($\epsilon$, purple) over the selected flight segment. Horizontal dashed lines indicate the corresponding time-averaged values, computed following \cite{chen2025optimal}.}
    \label{fig:S_fig05_1}
\end{figure}

\end{CJK}
\end{document}